\documentclass[preprint2]{aastex}

\newcommand{\neii}{[Ne~{\sc ii}]}

\newcommand{\hi}{H~{\sc i}}

\shorttitle{Gas in Brown Dwarf Disks}
\shortauthors{Pascucci et al.}

\begin{document}


\title{The Atomic and Molecular Content of Disks Around Very Low-mass Stars and Brown Dwarfs}


\author{I. Pascucci}
\affil{Lunar and Planetary Laboratory, The University of Arizona, Tucson, AZ 85721, USA}
\email{pascucci@lpl.arizona.edu}

\author{G. Herczeg}
\affil{Kavli Institute for Astronomy and Astrophysics, Peking University, Beijing, 100871, PR China}

\author{J. S. Carr}
\affil{Naval Research Laboratory, Code 7211, Washington, DC 20375, USA}

\and

\author{S. Bruderer}
\affil{Max Planck Institute for Extraterrestrial Physics, Giessenbachstrasse 1, 85748 Garching, Germany}



\begin{abstract}
There is growing observational evidence that disk evolution is stellar-mass dependent. Here, we show that these dependencies extend to the atomic and molecular content of disk atmospheres.
We analyze a unique dataset of high-resolution {\it Spitzer}/IRS spectra from 8 very low-mass star and brown dwarf disks. We report the first detections of Ne$^+$, H$_2$, CO$_2$, and tentative detections of H$_2$O toward these faint and low-mass disks. Two of our \neii{} 12.81\,\micron{} emission lines likely trace the hot ($\ge$5,000\,K) disk surface irradiated by X-ray photons from the central stellar/sub-stellar object. The H$_2$ S(2) and S(1) fluxes are consistent with arising below the fully or partially ionized surface traced by the \neii{} emission, in gas at $\sim$600\,K. We confirm the higher C$_2$H$_2$/HCN flux and column density ratio in brown dwarf disks previously noted from low-resolution IRS spectra. Our high-resolution spectra also show that the HCN/H$_2$O fluxes of brown dwarf disks are on average higher than those of T~Tauri disks. Our LTE modeling hints that this difference extends to column density ratios if H$_2$O lines trace warm $\ge 600$\,K disk gas. These trends suggest that the inner regions of brown dwarf disks have a lower O/C ratio than those of T~Tauri disks which may result from a more efficient formation of non-migrating icy planetesimals. A O/C=1, as inferred from our analysis, would have profound implications on the bulk composition of rocky planets that can form around very low-mass stars and brown dwarfs.
\end{abstract}


\keywords{circumstellar matter âmolecular processes â planetary systems: formation â planetary systems:
protoplanetary disks â stars: low-mass, brown dwarfs â stars: pre-main sequence}



\section{Introduction}

The mass and lifetime of gas in protoplanetary disks affect the formation and evolution of both giant and terrestrial planets. The connection to giant planets is perhaps the most obvious: giant planets must accrete most of their atmospheres from nebular gas hence the gas disk lifetime constrains their formation time (e.g., \citealt{lissauer07}). 
Once formed, giant planets can also migrate, as evinced by the detection of hot Jupiters around nearby stars (e.g., \citealt{howard12}). This migration can only occur in the presence of disk gas and the migration timescale strongly depends on the gas disk mass \citep{papa06}. 
In addition, giant planet atmospheres reflect the disk gas-phase abundance at their formation site, if contamination from planetesimal accretion is minimal. \cite{oberg11} discuss how the C$/$O ratio in hot Jupiters depends on the disk chemistry and the different snow lines of major carbon- and oxygen-bearing species and can be used to infer at which radial distances hot Jupiters formed.
Terrestrial planets are thought to form later than giant planets, likely in a gas-poor environment (e.g., \citealt{naga07}). However, residual gas might help circularizing their orbits \citep{kominami02} and determine their water content by transporting inward icy solids that can accrete onto forming terrestrial planets (e.g., \citealt{ciesla05}). Similarly to water, C is not accreted in in-situ formation models of Earth and may be delivered by accretion of icy bodies formed beyond the snowline \citep{bond10,lee10}.

In spite of its relevance to planet formation and the architecture of planetary systems, the amount, composition, and evolution of gaseous disks is poorly constrained. Mid-infrared observations with the {\it Spitzer Space Telescope} (hereafter, {\it Spitzer}) are enabling for the first time the study of disk gas atmospheres at radial distances where terrestrial and giant planets are expected to form. 

The first detailed analysis of high-resolution ($R\sim700$) {\it Spitzer} spectra from three accreting sun-like stars (hereafter, T~Tauri stars) revealed 
many rotational transitions from H$_2$O vapor and OH as well as ro-vibrational emission bands from simple organic molecules such as C$_2$H$_2$ and HCN \citep{carr08,salyk08}. Similar analysis extended to larger samples suggest that H$_2$O is abundant in the inner few AU of T~Tauri disks and traces a limited range of temperatures and column densities \citep{carr11,salyk11}. This limited range of excitation conditions is revealed by the similar relative strength of H$_2$O lines measured from star to star. In contrast, the relative strength of different organic molecules, and their strength with respect to water, present a much broader range. \cite{carr11} propose that the range in HCN/H$_2$O strength, in particular, results from different C/O ratios in the inner disk as a result of the migration history of icy planetesimals (see also \citealt{najita13}). This explanation hints that the composition of the gas disk atmosphere reflects radial transport in the disk midplane and perhaps planet formation timescales. On the other hand, the relative changes of H$_2$O and OH in the {\it Spitzer} spectra of the strongly variable T~Tauri star EX~Lupi points to stellar UV radiation as a key parameter for the disk surface chemistry \citep{banzatti12}. 
Similarly, stellar UV irradiation has been invoked to explain the lack of H$_2$O lines in {\it Spitzer} spectra of Herbig Ae/Be stars, which contrasts with two-thirds of T~Tauri disks having detectable water emission (\citealt{ponto10}, see also \citealt{fedele11} for the same result obtained by analyzing near-infrared molecular lines). 
Thus, both in-situ thermo-chemical processes and dynamical transport of icy planetesimals in the midplane may alter the composition of disk atmospheres.

\cite{pascucci09} have extended the study of gas disk atmospheres to very low-mass stars and brown dwarfs (hereafter, BD) using low-resolution ($R\sim120$) {\it Spitzer} spectra. They report the first detections of C$_2$H$_2$ bands in BD disks, lower HCN/C$_2$H$_2$ ratios but larger grain sizes than in T~Tauri disk atmospheres.
Motivated by these differences as a function of spectral type, here we analyze in detail the only high S/N spectra of very low-mass star and BD disks acquired with the high-resolution mode of the {\it Spitzer} IRS spectrograph \citep{houck04}. The sample comprises eight objects with spectral types (SpTy) ranging from M3 to M7.5, half of them have SpTy later than $\sim$M6 and are treated here as BDs (see e.g., \citealt{luhman07} for a discussion of the hydrogen-burning mass limit for young objects). The sources are
 located in the $\sim$1\,Myr-old Taurus \citep{kenyon08} and $\sim$10\,Myr-old Upper~Sco \citep{pecaut12} star-forming regions (see Table~\ref{table:propt}). Our sample is biased against sources with strong jets and may be slightly biased toward older objects. We report the first detections of \neii{} and H$_2$ pure rotational lines toward BDs (Sects.~\ref{sect:atomic} and \ref{sect:molecules}) and show that X-rays likely dominate over EUV in heating and ionizing the disk surface of these very low-mass objects (Sects.~\ref{discuss:neii} and \ref{discuss:warmh2}). We confirm the lower HCN/C$_2$H$_2$ abundance in BD disks with respect to T~Tauri disks (Sects.~\ref{sect:molecules} and ~\ref{sect:let}) previously reported by \citet{pascucci09}. We also report the first detections of H$_2$O lines in BD disks (Sect.~\ref{sect:molecules}) but observe an overall depletion of water vapor in comparison to other simple organic molecules. In Sect.~\ref{discussion:highC} we discuss the possibility that the high C$_2$H$_2$/HCN and HNC/H$_2$O ratios reflect an enhanced carbon chemistry in the inner regions of very low-mass star and BD disks.

\begin{deluxetable}{l c c c c c c c c c c}
\tabletypesize{\scriptsize}
\tablecaption{Summary of source properties\label{table:propt}}
\tablewidth{0pt}
\tablehead{
\colhead{ID} &\colhead{2MASS~J} & \colhead{Other Name} & \colhead{SpTy} &\colhead{d} & \colhead{mass} &\colhead{Log(L$_{\rm bol}$)}&
\colhead{Log(M$_{\rm acc}$)} & \colhead{Log(L$_{\rm X}$)} & \colhead{Log(L$_{\rm FUV}$)} & \colhead{Ref.} \\ 
\colhead{} &\colhead{} & \colhead{} & \colhead{} & \colhead{(pc)} & \colhead{(M$_\odot$)} &\colhead{(L$_\odot$)} &
\colhead{(M$_\odot$/yr)} & \colhead{(erg/s)}& \colhead{(L$_\odot$)}& \colhead{} }
\startdata
1 & 04381486+2611399 &		& M7.25  & 140 &0.07& -2.3 & -10.8 & -- & -- & 1,2,3,4  \\
2 & 04390163+2336029 & [SCH2006b] J0439016+2336030 &  M6 & 140 &0.08& -1.0 & --  & -- & -5.25 & 1,2,4,5,6   \\
3 & 04390396+2544264 &		& M7.25  & 140 &0.05& -1.7 & -11.2  & -- & -- & 1,2,3    \\
4 & 04442713+2512164	&		& M7.25  & 140 &0.05& -1.3 & -- &  27.6-28.6 & -4.93 & 1,2,4,5,6,7 \\  
5 & 05180285+2327127 & [SCH2006b] J0518028+2327126 & M5.5 &  140 &0.1& 0.02& -- & -- & -3.67& 2,5,6 \\
6 & 15582981-2310077	& UScoCTIO~33 &  M4.5  & 145 &0.16& -1.5 & -9.2 & -- & -- & 8,9   \\
7 & 16053215-1933159	&		&  M4.5 &  145 &0.16& -1.4 & -9.1  & -- & -- & 8,9  \\
8 & 16222160-2217307 & [SCH2006] J16222156-22173094 &  M5 & 145  &0.11 & -2.0 & -- & -- & -- 	& 8,10  \\ 
\enddata
\tablerefs{
(1) Luhman et al. 2010; 
(2) Kenyon et al. 2008;
(3) Muzerolle et al. 2005;
(4) Rebull et al. 2010;
(5) Herczeg \& Hillenbrand 2008;
(6) Yang et al. 2012;
(7) Guedel et al. 2007;
(8) de Zeeuw et al. 1999;
(9) Herczeg et al. in prep;
(10) Slesnick et al. 2006
}
\end{deluxetable}

\begin{deluxetable}{l c c c c c c}
\tabletypesize{\scriptsize}
\tablecaption{Log of the observations\label{table:log}}
\tablewidth{0pt}
\tablehead{
\colhead{ID} &\colhead{2MASS~J} &\colhead{AOR Key} & \colhead{PID} & \colhead{SH} & \colhead{SL} & \colhead{LL}  \\ 
\colhead{} &\colhead{} &\colhead{} & \colhead{} &\colhead{(time\tablenotemark{*}$\times$ ncycles)} & \colhead{(time\tablenotemark{*}$\times$ ncycles)} & \colhead{(time\tablenotemark{*}$\times$ ncycles)} }
\startdata
1 & 04381486+2611399 &      26924288,12705792 & 50799,2 & 120$\times$10 & 60$\times$12 & --   \\
2 & 04390163+2336029 & 26925824 & 50799& 120$\times$10 & 14$\times$2 & 14$\times$2 \\
3 & 04390396+2544264 &      26924800,18353664 & 50799,30540 & 120$\times$8  & 60$\times$1 & 30$\times$8 	 \\
4 & 04442713+2512164 &      26925312,12708608 & 50799,2 & 120$\times$8  & 14$\times$2 & 30$\times$2  \\  
5 & 05180285+2327127 & 26926336 & 50799& 120$\times$8  & 14$\times$2 & 14$\times$3    \\
6 & 15582981-2310077 & 26928384 & 50799& 120$\times$8  & 14$\times$3 & 14$\times$3     \\
7 & 16053215-1933159 &      26927872 & 50799& 120$\times$8  & 14$\times$3 & 14$\times$4 	\\
8 & 16222160-2217307 & 26926848 & 50799& 120$\times$10 & 14$\times$2 & 14$\times$2  \\ 
\enddata
\tablecomments{SH spectra are all from PID~50799. 2MASS~04381486+2611399 is the only source without a LL spectrum.}
\tablenotetext{*}{Exposure times are in seconds}
\end{deluxetable}

\section{Observations and Data Reduction}
Observations were carried out with the {\it Spitzer} IRS  short-high (SH) module in May 2009 as part of the program PID~50799 (PI, G. Herczeg). The SH module covers the wavelength range $\sim$10--19\,\micron{} with a resolving power of approximately 700. Lower resolution spectra were also acquired within the same program using the short-low (SL: 5.2--14.5\,\micron) and long-low modules (LL: 14--38\,\micron) for all but three sources (objects \#1, 3, and 4). For these three sources we retrieved and reduced archival data from other programs (see Table~\ref{table:log}). 2MASS~J04381486+2611399 (object \#1) is the only BD that was not observed with the LL module, hence its low-resolution spectrum extends only out to 14.5\,\micron . Table~\ref{table:log} provides the observational log.

\begin{figure*}
\includegraphics[angle=90,width=\textwidth]{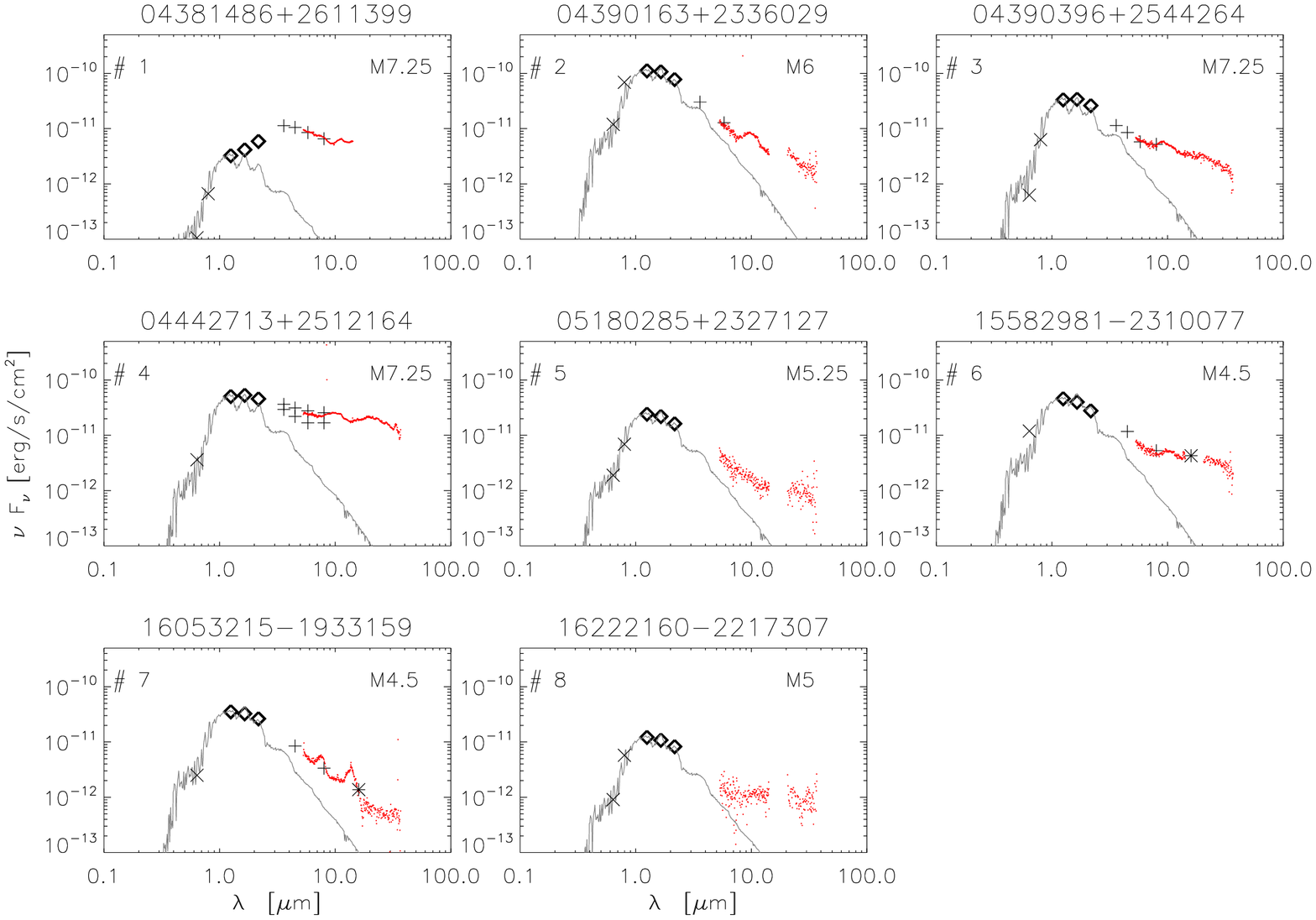}
\caption{Spectral energy distributions of our targets. We plot R and I (X),  
2MASS J, H, and K (diamonds), IRAC 3.6, 4.5, 5.8, and 8\,\micron{} (+), and 
MIPS 24\,\micron{} (asterisk) fluxes. Low-resolution {\it Spitzer}/IRS 
spectra are shown with dot red symbols. NEXTGEN model atmospheres are
plotted with grey lines.  Objects \#1 and 4 have the largest infrared excess among our targets.}
\label{fig:seds}
\end{figure*}

The low-resolution spectra were reduced using the data reduction scripts developed by the "Formation and Evolution of Planetary Systems" Spitzer Legacy team \citep{meyer06}. We refer to \citet{bouwman08} for details about the data reduction. These spectra are primarily used to verify the flux calibration and spectral shape of the high-resolution spectra. They are also used, in combination with published photometry, to characterize the spectral energy distribution (SED) of our eight sources (see Fig.~\ref{fig:seds}).
In what follows, we focus on describing the observational strategy and reduction of the high-resolution data.

Targets were acquired with the visual Pointing Calibration and Reference Sensor and placed on two positions along the slit (at 1/3 and 2/3 of the slit length) following the standard IRS staring mode technique. Sky exposures were obtained right after the on-source exposures at 60$''$ north and south of each target. These exposures are used to remove the infrared background emission and to reduce the number of {\it rogue} pixels. The data reduction was carried out with {\it idl} custom routines following the steps outlined in \citet{pascucci06} and the implementations on the noise statistics and bad pixels identification developed by \citet{carr11}. After image combination, background subtraction, and pixel correction, we extracted one-dimensional spectra with the full aperture extraction routine in SMART \citep{higdon04}. We processed in the same way ten high S/N spectra of the calibrator $\xi$ Dra to create one-dimensional spectral response functions (for each order and at each nod position) from the known stellar model atmosphere\footnote{http://irsa.ipac.caltech.edu/data/SPITZER/docs/irs
/calibrationfiles/decinmodels/}. We divided the source spectra by the spectral response functions and combined spectral orders using appropriate weighting in the overlapping regions. Finally, we averaged the flux-calibrated spectra at the two nod positions and took their absolute difference divided by two to estimate the noise in the average spectra.

\begin{figure*}
\includegraphics[angle=90,width=\textwidth]{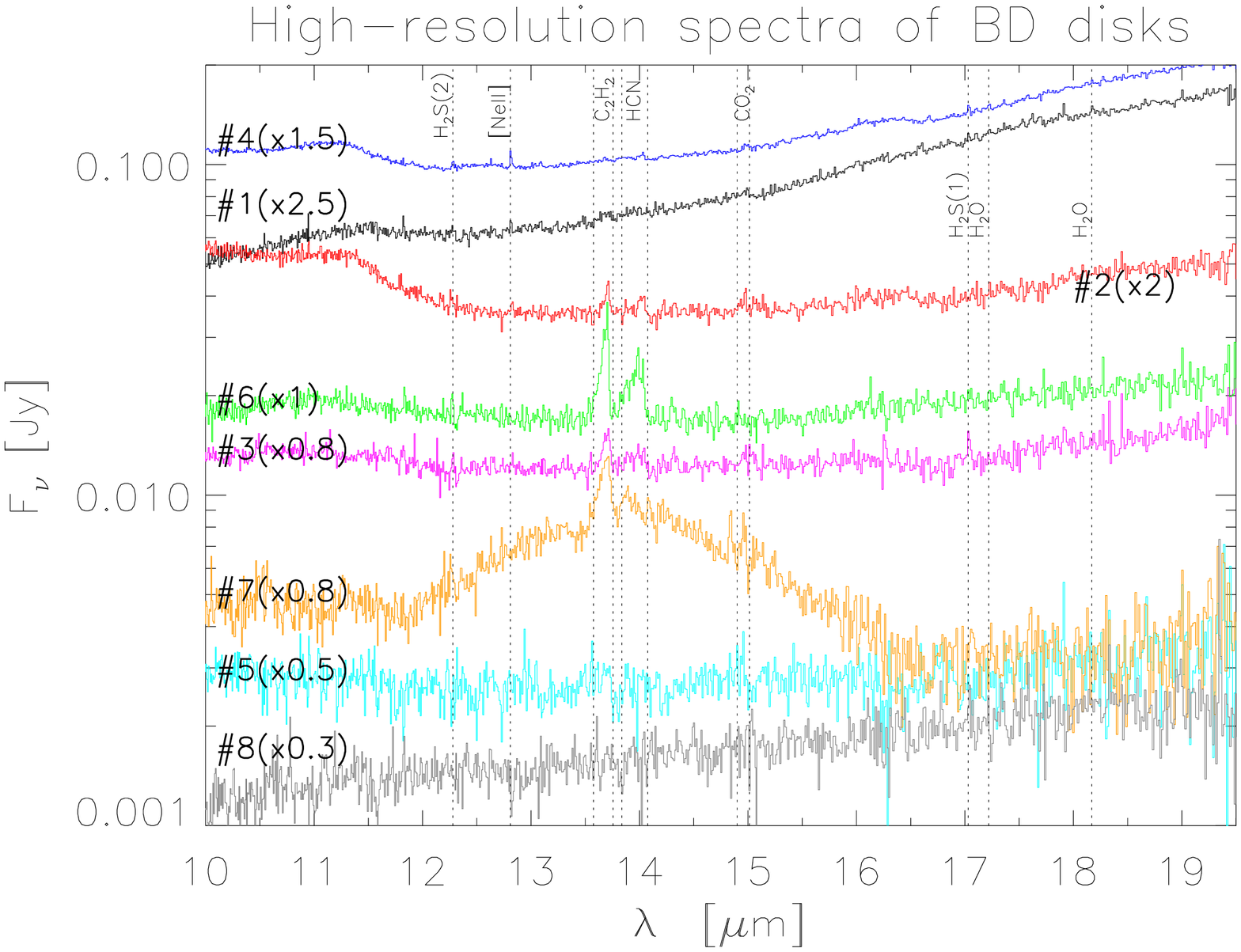}
\caption{High-resolution {\it Spitzer} spectra of our targets. Bands and emission lines discussed in the text are marked with dotted lines}. Spectra have been
scaled by the factor in parenthesis to better separate them. The broad emission between 12 and 16\,\micron{} in object \#7 is further discussed in Appendix~\ref{obj7}.
\label{fig:highresspk}
\end{figure*}

The reduced SH spectra are shown in Fig.~\ref{fig:highresspk}. For all objects except \#4, the high-resolution spectra match in flux and shape with the low-resolution spectra, within the 10\% photometric accuracy of the IRS high-resolution mode. The low-resolution spectrum of object  \#4 is overall $\sim$20\% higher than the high-resolution spectrum. Because the low- and high-resolution spectra are not contemporaneous (see Table~\ref{table:log}) and source \#4 showed a $\sim$30\% change in its {\it Spitzer}/IRAC fluxes in one year (Fig.~\ref{fig:seds} and \citealt{luhman10}), we attribute the difference between the high- and low-resolution spectra to intrinsic source variability.

\section{Observed Infrared Lines}\label{sect:infraredspectra}
Before presenting the atomic and molecular lines detected in our sample,
we compute additional stellar (BD) parameters that will be useful in interpreting the observations. We also briefly discuss the SEDs of our very low-mass stars and BDs.

Table~\ref{table:propt} provides literature SpTy, distances, masses, and bolometric luminosities for all our sources. Mass accretion rates (M$_{\rm acc}$) are available for four sources while three sources have FUV luminosities (L$_{\rm FUV}$) from the literature, see Table~\ref{table:propt}. Object \#8 (16222160-2217307) is the only one for which there are no M$_{\rm acc}$ or L$_{\rm FUV}$ estimates. Sources with L$_{\rm FUV}$ are from the far-UV atlas of \citet{yang12} and have also accretion luminosities (L$_{\rm acc}$) reported in the same paper. For these three sources we compute M$_{\rm acc}$ using the standard relationship with L$_{\rm acc}$ (e.g. eq. 8 in \citealt{gullbring98}), fixing the disk inner radius to 5 times the stellar radius. The stellar radius is computed using the Stefan-Boltzmann relation with the bolometric luminosity given in Table~\ref{table:propt} and the stellar effective temperature from the source SpTy (also in Table~\ref{table:propt}) and the temperature scale of \citet{luhman03}. M$_{\rm acc}$ calculated this way are reported in Table~\ref{table:inferred}. For the four sources with 
literature M$_{\rm acc}$ we use the relationship mentioned above to compute L$_{\rm acc}$ and then convert 
L$_{\rm acc}$ into L$_{\rm FUV}$ following \citet{yang12}. L$_{\rm FUV}$ computed this way are reported in Table~\ref{table:inferred}. Finally, only BD \#4 has a measured X-ray luminosity (see Table~\ref{table:propt}). We estimate L$_{\rm X}$ for all other sources using the empirical relation between  X-ray luminosity and stellar mass found by \citet{guedel07} and list the estimated L$_{\rm X}$ in Table~\ref{table:inferred}.

\begin{deluxetable}{l c c c c}
\tabletypesize{\scriptsize}
\tablecaption{Estimated stellar and brown dwarf properties\label{table:inferred}}
\tablewidth{0pt}
\tablehead{
\colhead{ID} &\colhead{2MASS~J} &\colhead{Log(M$_{\rm acc})$} &\colhead{Log(L$_{\rm X})$} & \colhead{Log(L$_{\rm FUV}$)}  \\ 
\colhead{} &\colhead{} &\colhead{(M$_\odot$/yr)} & \colhead{(erg/s)} &\colhead{(L$_\odot$)}
}
\startdata
1 & 04381486+2611399 & \tablenotemark{*} &28.2 & -4.86      \\
2 & 04390163+2336029 & -9.8 &28.3 & \tablenotemark{*}\\
3 & 04390396+2544264 & \tablenotemark{*} & 28.0 & -5.81  	 \\
4 & 04442713+2512164 &  -11.0 & \tablenotemark{*} &    \tablenotemark{*}  \\  
5 & 05180285+2327127 &  -9.3 &28.5 & \tablenotemark{*}  \\
6 & 15582981-2310077 &  \tablenotemark{*} &28.8 & -3.67   \\
7 & 16053215-1933159 &  \tablenotemark{*} &28.8 & -3.59    	\\
8 & 16222160-2217307 & --  &28.5 & --\\ 
\enddata
\tablenotetext{*}{See Table~\ref{table:propt} for values reported in the literature}
\end{deluxetable}

The SEDs of our very low-mass stars and BDs are shown in Fig.~\ref{fig:seds}. They include optical and infrared photometry (R, I, J, H, K,  {\it Spitzer}/IRAC and MIPS) and our low-resolution {\it Spitzer}/IRS spectra. We also show the NEXTGEN \citep{hauschildt99} model atmosphere (scaled to the source J-band flux) appropriate to each source (source SpTy in Table~\ref{table:propt} and temperature scale in \citealt{luhman03}).
Six out of eight sources have moderate infrared excess emission, with observed fluxes at 8\,\micron{} that are between 3 and 7 times the photospheric fluxes. BDs \#1 and \#4 have fluxes that are more than an order of magnitude larger than their photospheric fluxes.
BD \#1 is also the least luminous at optical wavelengths in agreement with being surrounded by an almost edge-on disk \citep{luhman07}. The IRS low-resolution spectrum of object \#7 shows two prominent and broad bands between 7-8 and 12-16\,\micron{} (Fig.~\ref{fig:seds}) that appear to be emission features. The feature at longer wavelengths is also clearly seen in the high-resolution spectrum (Fig.~\ref{fig:highresspk}). These bands are not spatially extended beyond the relatively narrow slit of the low-resolution spectrograph ($\sim 500$\,AU projected width), hence must be associated to the star/disk system. In Appendix~\ref{obj7} we speculate on the possible origin of these features.


\subsection{Atomic/Ionic lines} \label{sect:atomic}
\begin{figure*}
\includegraphics[angle=90,scale=.70]{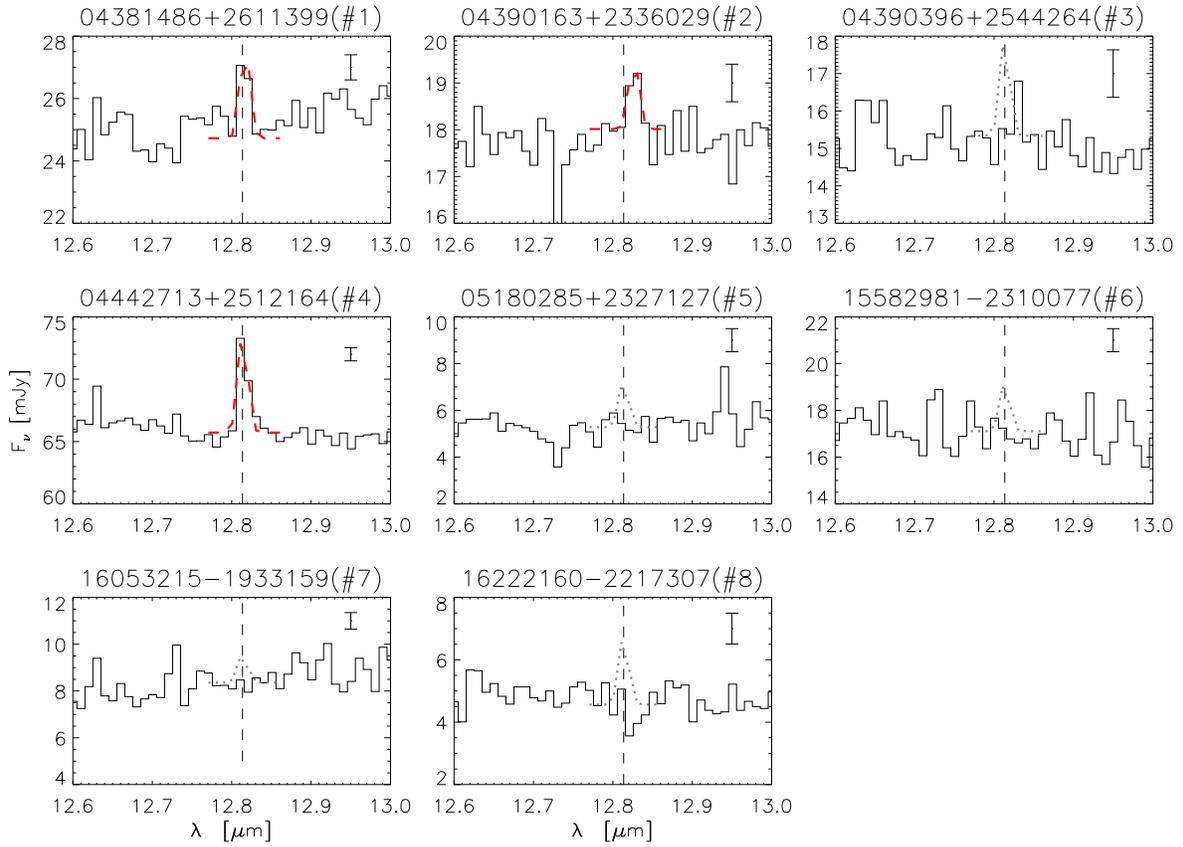}
\caption{Expanded view of the wavelength region around the \neii{} line.
On top of the continuum emission we overplot the best Gaussian fits to the data
(red dashed lines) or the hypothetical 3$\sigma$ upper limits
(grey dotted lines) in case of non-detections. The errorbar provides the 1$\sigma$ uncertainty based on the rms on continuum-subtracted spectra, see text.
}
\label{fig:neiidet}
\end{figure*}

The strongest atomic/ionic lines detected in {\it Spitzer} spectra of T~Tauri disks are the \hi{}(7-6) at 12.37\,\micron{} and the \neii{} at 12.81\,\micron{} (e.g., \citealt{pascucci07}, \citealt{lahuis07}). We have searched for these emission lines in our sample and report the first \neii{} detections toward BD disks (see Fig.~\ref{fig:neiidet}).

To estimate line fluxes and upper limits we followed the procedure outlined in \citet{pascucci07}. 
In brief, when a \neii{} line is detected, we fit a Gaussian plus a first-order polynomial to the spectral region within $\pm$0.05\,\micron{} of the line (10 pixels). In case of non-detections, we fit the same wavelength region with a first-order polynomial and compute a 3$\sigma$ upper limit from the rms of the continuum-subtracted spectrum. The 1$\sigma$ uncertainties on detected lines are also computed from the rms of continuum-subtracted spectra in a region just outside the line.
Our results are summarized in Table~\ref{table:neiih2}. The brightest \neii{} emission is from 
BD \#4 with a line flux of $\sim 23 \times 10^{-16}$\,erg\,s$^{-1}$\,cm$^{-2}$. 
Line fluxes from BD \#1 and \#2 are $\sim 3$ and 5 times weaker respectively. 
We also note that the emission toward source \#2 is only a tentative detection (slightly above 2$\sigma$) and should be confirmed via future observations with JWST.  For the other five targets we estimate 3$\sigma$ upper limits in the \neii{} line that range between (4--9)$\times 10^{-16}$\,erg\,s$^{-1}$\,cm$^{-2}$. Similar upper limits are found for the HI(7-6) line which is never detected in our sample.
We will discuss in Sect.~\ref{discuss:neii} the likely origin of the \neii{} emission toward very low-mass stars and BDs.

\subsection{Molecular transitions} \label{sect:molecules}
The {\it Spitzer} IRS SH module covers many rotational transitions from H$_2$O and OH as well as rovibrational bands from simple molecules such as  C$_2$H$_2$, HCN, and CO$_2$ (Q-branches at 13.7, 14.0, and 14.97\,\micron, respectively). These molecules are frequently detected toward T~Tauri stars with disks and likely trace the warm disk atmosphere (e.g., \citealt{ponto10}). 
The H$_2$ infrared lines have been also detected
toward T Tauri disks \citep{lahuis07,carr11}, though
some detections may be tracing outflows or cloud
emission rather than the disk itself \citep{lahuis07}.

\begin{figure*}
\centering
\includegraphics[width=4in]{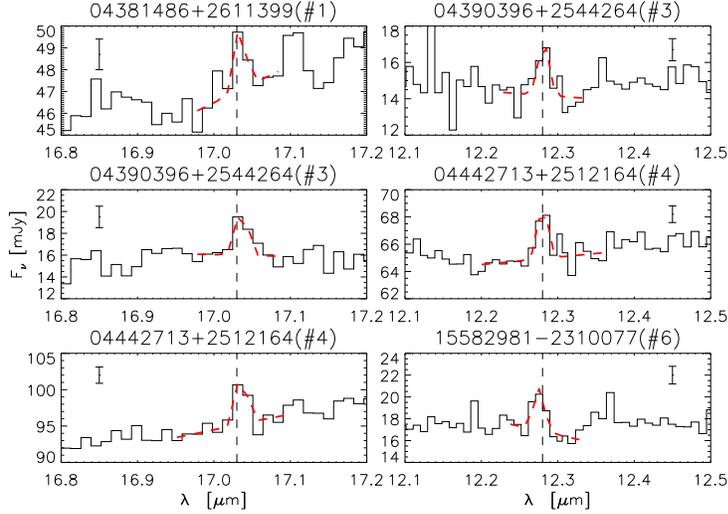}
\caption{Expanded view of the wavelength region around the H$_2$ lines.
We only show spectra where either the H$_2$ S(2) at 12.28\,\micron{} or the S(1) line at 17.03\,\micron{} have been detected. 
On top of the continuum emission we overplot the best Gaussian fits to the data
(red dashed lines). The emission around 17.1\,\micron{} toward 2MASS~J04381486+2611399 (\#1) may be from H$_2$O (see Fig.~\ref{fig:h2odet}). The errorbar provides the 1$\sigma$ uncertainty based on the rms on continuum-subtracted spectra, see text.}
\label{fig:h2det}
\end{figure*}

Here, we report the first detections of H$_2$ pure rotational lines toward BD disks (see Fig.~\ref{fig:h2det}). The procedure to measure line fluxes and upper limits is identical to that used for the \neii{} lines (Sect.~\ref{sect:atomic} and Table~\ref{table:neiih2}). Firm H$_2$ detections ($\ge$3$\sigma$) are only available for sources \#3, 4, and 6 in the H$_2$\,S(2) at 12.28\,\micron{} and for source \#4 in the H$_2$\,S(1) at 17.03\,\micron. Tentative detections (above 2$\sigma$) are also reported for \#1 and 3 in the H$_2$\,S(1) transition, see Fig.~\ref{fig:h2det} and Table~\ref{table:neiih2}. Objects \#4 is the only one where both H$_2$ transitions are firmly detected and point to similar line fluxes in both transitions. When lines are not detected 3$\sigma$ upper limits are found to be $\sim(4-10)\times 10^{-16}$\,erg\,s$^{-1}$\,cm$^{-2}$. We will further discuss these detections and upper limits in Sect.~\ref{discuss:warmh2}.



\begin{deluxetable}{l c c c c}
\tabletypesize{\scriptsize}
\tablecaption{Line fluxes and 3$\sigma$ upper limits for \neii{} and H$_2$ transitions.\label{table:neiih2}}
\tablewidth{0pt}
\tablehead{
\colhead{ID} &\colhead{2MASS~J} &\colhead{\neii} & \colhead{H$_2$ S(2)} & \colhead{H$_2$ S(1)}  
}
\startdata
1 & 04381486+2611399 & 8.5$\pm$1.9 & $<$10 & 5.6$\pm$2.5 \\
2 & 04390163+2336029 & 4.3$\pm$1.9 &  $<15$ & $<6$  \\
3 & 04390396+2544264 & $<$9 & 10$\pm$3  & 8.5$\pm$3.5  \\
4 & 04442713+2512164 & 22.6$\pm$2.5 & 13$\pm$3  & 13$\pm$4 \\
5 & 05180285+2327127 & $<$7 &  $<12$&  $<6$ \\
6 & 15582981-2310077 & $<$7 &13$\pm$4  & $<10$ \\
7 & 16053215-1933159 & $<$5 & $<$10 & $<6$  \\
8 & 16222160-2217307 & $<$7 &  $<6$& $<$9  \\
\enddata
\tablecomments{Fluxes are in units of 10$^{-16}$\,erg\,s$^{-1}$\,cm$^{-2}$.}
\end{deluxetable}

\begin{figure*}
\includegraphics[angle=90,scale=.70]{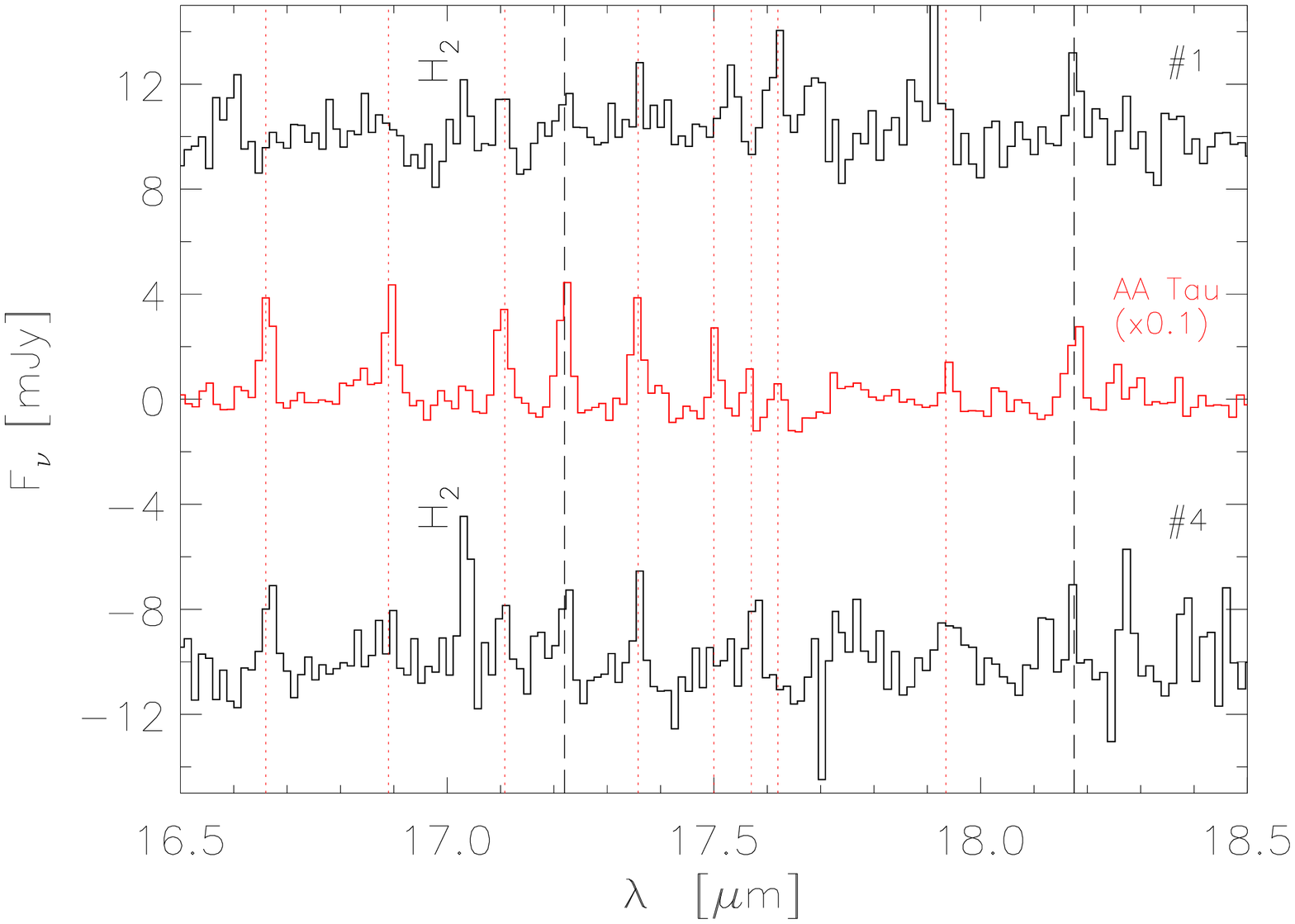}
\caption{Continuum-subtracted spectra (black lines) of the two BDs toward which we have tentative detections of water lines. The continuum-subtracted spectrum of the T~Tauri star AA~Tau \citep{carr08} is plotted in the middle of the figure in red. A constant shift has been applied to BD spectra while the spectrum of AA~Tau has been scaled by the factor in parenthesis to facilitate the comparison with the BD spectra. Black vertical dashed lines mark the 17.22 and 18.17\,\micron{} H$_2$O transitions for which we report fluxes or upper limits in Table~\ref{table:organic_water}. Red vertical dotted lines mark other water transitions that are detected in AA~Tau.
}
\label{fig:h2odet}
\end{figure*}

We also report detections of the C$_2$H$_2$ and HCN rovibrational bands toward four disks (objects \#2, 3, 6, and 7, see later the continuum-subtracted spectra in Fig.~\ref{fig:moldet}, the latter HCN detection is very weak). In all objects the C$_2$H$_2$ band emission appears stronger than the HCN emission. This pattern was already recognized in low-resolution {\it Spitzer} spectra from 19 very low-mass stars/BDs belonging to the Cha~I star-forming region \citep{pascucci09} and is opposite to what is seen in T~Tauri disks \citep{pascucci09,carr11}. The five times higher resolution of the spectra presented here enables us to detect fainter emission and model the shape of the band emission (see Sect.~\ref{sect:let}). We also present the first CO$_2$ detections toward BD disks (objects \#2, 3, and 7) and tentative detections of H$_2$O lines in the spectra of objects \#1 and \#4 (see Fig.~\ref{fig:h2odet}). We do not have any tentative or firm detection of  OH lines in our spectra. 

\begin{figure*}
\centering
\includegraphics[angle=0,scale=.70]{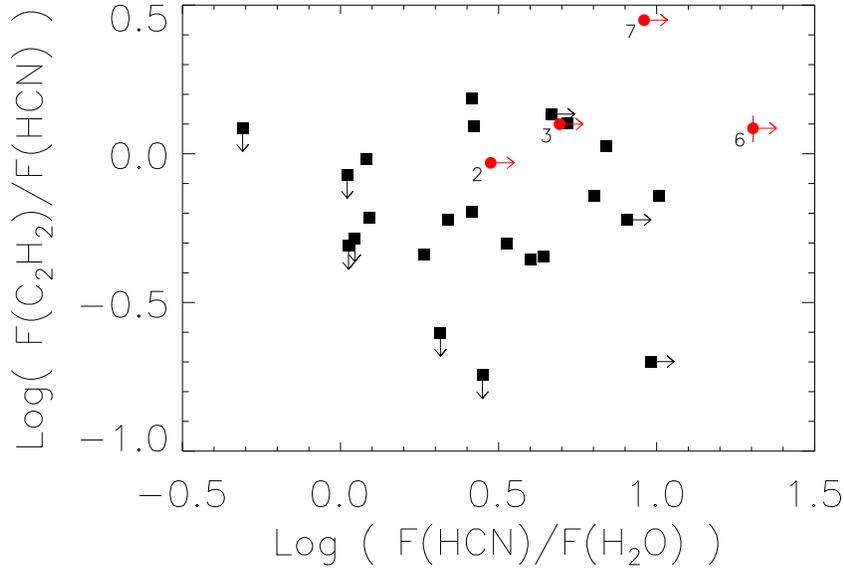}
\caption{Log-log plot of molecular flux ratios (see also Table~\ref{table:organic_water}). For comparison with other studies we use the H$_2$O transition at 17.22\,\micron . Red circles are very low-mass star and BD disks from this study. Black squares are T~Tauri disks from \citet{ponto10} and \citet{salyk11}. Our sample occupies the upper right portion of the plot: C$_2$H$_2$ fluxes $\ge$ HCN fluxes and HCN fluxes more than $\sim$3 times the H$_2$O fluxes at 17.22\,\micron .}
\label{fig:fluxratios}
\end{figure*}

To quantify the strength of the molecular emission, we measure fluxes in the H$_2$O 17.22 and 18.17\,\micron{} lines\footnote{We do not have any detection above 3$\sigma$ in the 15.17\micron{} H$_2$O line which is often detected in T~Tauri disks, see \cite{ponto10}, therefore we do to include it in our discussion} and in the organic molecular bands using a procedure similar to that outlined in \cite{salyk11}.
First, we fit a linear continuum using pixels shortward and longward of the emission. Then, we subtract the continuum emission and integrate the flux within wavelength regions that cover the full emission (see Note to Table~\ref{table:organic_water}). Line flux upper limits are calculated from the rms in the continuum-subtracted spectra using line widths appropriate to each transition: an unresolved line width for the H$_2$O rotational transitions, three times the resolution for the C$_2$H$_2$  and HCN bands, and two times the resolution for the CO$_2$ band. The 1$\sigma$ uncertainties on detected lines are also calculated from the rms in the continuum-subtracted spectra. Table~\ref{table:organic_water} reports our line flux measurements and 3$\sigma$ upper limits when lines are not detected. The strongest band emission is typically from C$_2$H$_2$ followed by HCN and by CO$_2$. Water lines are the weakest and tentatively detected at the 2-3$\sigma$ level only toward the two sources with no emission from organic molecules (objects \#1 and \#4).  These tentative detections should be prime candidates for followup observations with JWST. Fig.~\ref{fig:fluxratios} compares molecular line flux ratios from our sample with those from T~Tauri disks \citep{ponto10,salyk11}. While there is some overlap between the two samples, very low-mass star and BD disks tend to occupy the upper right corner of the plot. The data in Fig.~\ref{fig:fluxratios} confirm the result from \citet{pascucci09}
that BD disks have larger C$_2$H$_2$/HCN flux ratios than the average
T~Tauri disk.
In addition, it shows that the HCN/H$_2$O (and C$_2$H$_2$/H$_2$O) flux ratios tend to be larger in BD than in T~Tauri disks. We will further discuss the significance of this result in Sect.~\ref{discussion:highC}.

\begin{deluxetable}{l c c c c c c}
\tabletypesize{\scriptsize}
\tablecaption{Line fluxes and 3$\sigma$ upper limits for organic molecules and water lines.\label{table:organic_water}}
\tablewidth{0pt}
\tablehead{
\colhead{ID} &\colhead{2MASS~J} &\colhead{C$_2$H$_2$} & \colhead{HCN} & \colhead{CO$_2$} &  
\colhead{H$_2$O(17.22\,\micron)} &  \colhead{H$_2$O(18.17\,\micron)}
}
\startdata
1 & 04381486+2611399 & $<$32 & $<$31 & $<$18  & 5$\pm$2 & 6$\pm$2 \\
2 & 04390163+2336029 & 33.8$\pm$9 & 36.2$\pm$9 & 14.5$\pm$5  & $<$12 & $<$7\\
3 & 04390396+2544264 & 56.0$\pm$11 & 44.5$\pm$10 & 9.6$\pm$4  & $<$9 & $<$17\\
4 & 04442713+2512164 & $<$23\tablenotemark{a} & $<$38\tablenotemark{a} & $<$24 & 8$\pm$3 & $<$6\\
5 & 05180285+2327127 & $<$25 & $<$24 & $<$12  & $<$7  & $<$7\\ 
6 & 15582981-2310077 & 205$\pm$14 & 169$\pm$13  & $<$21 & $<$8 & $<$19\\
7 & 16053215-1933159 & 106$\pm$11 & 37.6$\pm$11 & 11.1$\pm$4.8 & $<$4 & $<$7  \\ 
8 & 16222160-2217307 &  $<$21 & $<$20 &  $<$14 & $<$6  & $<$5\\
\enddata
\tablecomments{Fluxes are in units of 10$^{-16}$\,erg\,s$^{-1}$\,cm$^{-2}$. Integrations are performed over the following wavelength ranges: 
C$_2$H$_2$ (13.576-13.756\micron); HCN (13.837-14.075\micron); CO$_2$ (14.900-15.014\micron); H$_2$O (17.19-17.25\micron{} and 18.12-18.22\micron). 3$\sigma$ upper limits 
are computed for an unresolved line for the H$_2$O rotational line, over two times the resolution for CO$_2$, and over three times the resolution for 
C$_2$H$_2$ and HCN. This choice reflects the width of the rotational and rovibrational lines when detected.}
\tablenotetext{a}{HCN and to less extent C$_2$H$_2$ emission are clearly contaminated by water lines.}
\end{deluxetable}

\section{LTE modeling of H$_2$O and organic molecules}\label{sect:let}

Water and OH are commonly detected in the IR spectra of T~Tauri stars and their many transitions can contaminate emission lines/bands from other molecules (see e.g. \citealt{carr08}). This is not the case for our sample where weak H$_2$O lines are tentatively detected in  only two BD spectra, objects \#1 and 4, those that do not present emission from other organic molecules  (see Table~\ref{table:organic_water} and  Fig.~\ref{fig:h2odet}). Hence, when modeling the emission bands of organic molecules from objects \#2, 3, 6, and 7, we can safely ignore any contamination from water. However, we still model the water emission to derive H$_2$O column density upper limits based on the line flux upper limits reported in Table~\ref{table:organic_water}.


We compute synthetic spectra of H$_2$O and organic molecules adopting a plane parallel slab model of gas with a single temperature and column density. We assume the gas to be in local thermal equilibrium (LTE) and account for optical depth effects as in \citet{banzatti12}. Previous studies have shown that the LTE assumption can relatively well reproduce the line luminosities of most H$_2$O lines  detected in T~Tauri disks with only the higher energy level transitions hinting to a possible departure from LTE \citep{carr11,salyk11}. However, see \citet{meijerink09} for the importance of non-LTE effects in estimating column densities, especially in the case of H$_2$O whose mid-infrared transitions span four orders of magnitudes in critical density ($\sim 10^8 -10^{12}$\,cm$^{-3}$).
In our simple LTE model there are four input parameters to produce a synthetic spectrum: the gas temperature ($T$), the line-of-sight column density ($N$), the radius ($R$) of the projected emitting area, and the local line width. To allow a direct comparison with models of T~Tauri/Herbig Ae/Be disks we use as local line width that due only to thermal broadening, which is temperature- and molecule-dependent. 

\begin{deluxetable}{l c c c c c c c}
\tabletypesize{\scriptsize}
\tablecaption{Modeling results.\label{table:resmolfit}}
\tablewidth{0pt}
\tablehead{
\colhead{ID} & \colhead{2MASS~J} &\colhead{$T$} & \colhead{$R$} &\colhead{$N_{\rm C_2H_2}$} & \colhead{$N_{\rm HCN}$} & \colhead{$N_{\rm CO_2}$} &  \colhead{$N_{\rm H_2O}\tablenotemark{\Box}$} \\ 
\colhead{} & \colhead{} &\colhead{(K)} & \colhead{(AU)} &\colhead{($10^{16}$\,cm$^{-2}$)} &\colhead{($10^{16}$\,cm$^{-2}$)} & \colhead{($10^{16}$\,cm$^{-2}$)} & \colhead{($10^{16}$\,cm$^{-2}$)} }
\startdata
	 
2 & 04390163+2336029 & 240 & 0.29 & 15.8   & 18.1  & 9.5 & $<$ (1[+9],5) \\
3 & 04390396+2544264 & 240 & 0.26 & 50.1  & 11.7 & 5.6 & $<$  (1[+9],21) \\      
6 & 15582981-2310077 & 960 & 0.09 & 12.6  & 8.4 &  $<1$ & $<$ (65,2[+4]) \\
7 & 16053215-1933159\tablenotemark{*} & 960 &0.09 & 4.5 & 1.1 & 0.9	& $<$ (11,333) \\ 
\enddata
\tablecomments{For all sources except source \#7, $T$, $R$, and $N_{\rm C_2H_2}$ are best-fit values from modeling the C$_2$H$_2$ band. 
$N_{\rm HCN}$, $N_{\rm CO_2}$, and the first entry of $N_{\rm H_2O}$ are computed by fixing $T$ and $R$ to those found via the C$_2$H$_2$ modeling. The second entry of $N_{\rm H_2O}$ is obtained using the same $R$ as above but assuming a temperature of 600\,K. }
\tablenotetext{*}{For this source we cannot find a good fit to the C$_2$H$_2$ band, the emission is too broad. We have thus fixed the temperature and emitting area to those of source \#6 and varied the column density to reproduce the observed flux integrated over the emission bands and lines.}
\tablenotetext{\Box}{We used the 3$\sigma$ upper limits to the 17.22 and 18.17\,\micron{} H$_2$O lines reported in Table~\ref{table:organic_water} and give here the most stringent constraint on the column density.
The first entry is for the gas temperature listed in column three of this table while the second entry is for $T=600$\,K. Values in square brackets of the form [+b] stand for $10^{+b}$. Note that for objects  \#2 and 3 the water column density is essentially unconstrained for the low temperature case ($T=240$\,K).}
\end{deluxetable}

\begin{figure*}
\centering
\includegraphics[scale=0.7]{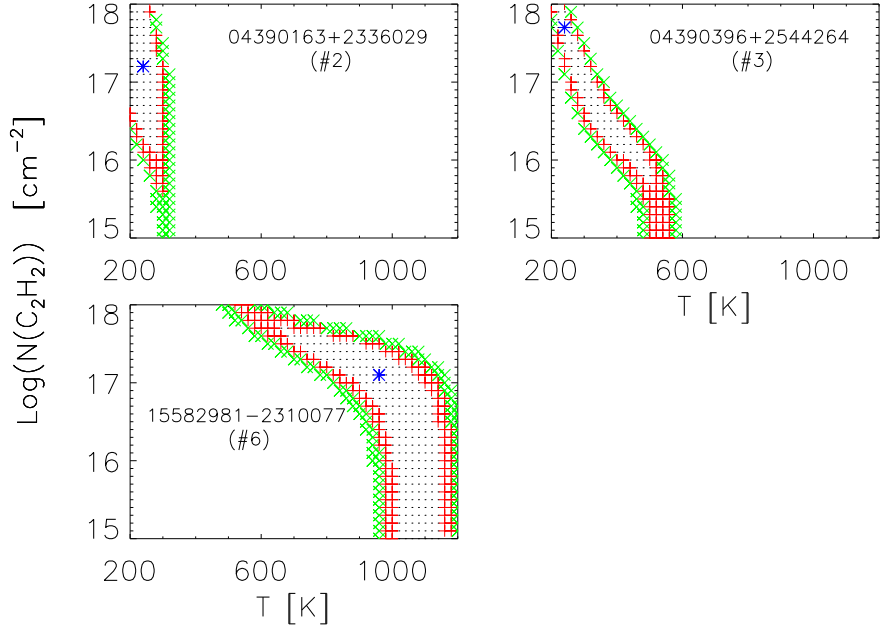}
\caption{Plots of $\chi^2$ as a function of temperature (T) and column density (N) for fits to the C$_2$H$_2$ Q-branch band emission.
A blue asterisk marks the bet-fit values, minimum $\chi^2$. Black dots, red crosses, and green 'X' mark the regions within 1$\sigma$, 2$\sigma$, and 3$\sigma$ respectively 
when assuming a gaussian $\chi^2$ distribution.}
\label{fig:contplot}
\end{figure*}

To compare the observed and synthetic spectra of C$_2$H$_2$, HCN, and CO$_2$, we first
estimate the continuum level on three wavelength regions, two outside and one in between the emission bands. We find that second-order polynomials can reproduce well the continuum except for object \#7 where the broad emission between 14 and 16\,\micron{}  
is best modeled with a fourth-order polynomial. Next, we subtract the continuum emission and use the LTE model described above to reproduce the continuum-subtracted band emission from each molecule. 
Because the C$_2$H$_2$ rovibrational band is typically the strongest, we first run a grid of models with temperatures varying between 50 and 1200\,K (in steps of 20\,K) and column densities between $10^{14}$ and $10^{18}$\,cm$^{-2}$ (in logarithmic steps of 0.1) to reproduce its profile. The radius of the projected emitting area is varied to match the integrated area below the emission band but it is kept smaller than 1\,AU. In other words, solutions with emitting radii extending beyond 1\,AU are not considered plausible, given the aforementioned considerations on the extension of the mid-infrared region in BD disks, and thus are discarded. We compute $\chi^2$ from the difference of the model and observed spectrum for each grid point to estimate the best-fit values and confidence intervals. Our results are summarized in Table~\ref{table:resmolfit} and Fig.~\ref{fig:contplot}. Only for source \#7 we could not find good fits within the explored parameter range. Its C$_2$H$_2$ band emission appears too broad even for the highest temperature in our grid (1,200\,K). There is also some extra emission around 13.85\,\micron{} which is not seen toward other sources and may be due to contamination from other molecules not considered in our model. For this source we assume the same gas temperature and emitting radius as for source \#6 and compute the $N_{\rm{C_2H_2}}$ that reproduces the observed flux integrated over the band.

Fig.~\ref{fig:contplot} shows the $\chi^2$ as a function of the gas temperature $T$ and column density $N_{\rm{C_2H_2}}$ for our fits to the C$_2$H$_2$ Q-branch band emission of sources \#2, 3, and 6.
With the assumption that the $\chi^2$ distribution is gaussian we also draw the 1$\sigma$, 2$\sigma$, and 3$\sigma$ confidence intervals. This figure is instructive because it shows that there is a relatively large range of plausible $T$ and $N_{\rm{C_2H_2}}$ (even within 1$\sigma$) and that these two input parameters are degenerate (higher $T$ is offset by lower $N$).
This same behavior has been noted in previous studies of T~Tauri disks \citep{carr11,salyk11}. If we consider the 1$\sigma$ intervals we see that the gas disk temperature of the Taurus BD \#2 can be as low as 200\,K and up to 280\,K with $N_{\rm{C_2H_2}}$ varying between 10$^{16-18}$\,cm$^{-2}$. The gas temperature of BD \#3, also in Taurus, can be as high as 500\,K and column densities can be as low as $5\times10^{15}$\,cm$^{-2}$ for these higher $T$. The disk surface of objects \#6 and \#7, which are in Upper~Sco, is definitively hotter than that of the Taurus BDs~\#2 and 3. Within the 1$\sigma$ interval, $T$ varies between 660 and 1140\,K and $N_{\rm{C_2H_2}}$ between $1-500\times10^{15}$\,cm$^{-2}$ for object \#6. While the average bolometric and X-ray luminosities of \#6 and 7 are only about a factor of 3 higher than the average values for BDs~\#2 and 3, their average FUV luminosity is more than a factor of 50 higher (see Tables~\ref{table:propt} and \ref{table:inferred}). The overall higher heating may be responsible for the hotter disk surface of \#6 and 7 we infer from our fits to the C$_2$H$_2$ rovibrational bands (see also Sect.~\ref{discussion:highC}).
We note that at temperatures as high as $\sim$1,000\,K  PAHs and C$_2$H$_2$ can be closely related and the destruction of PAHs may enhance the abundance of C$_2$H$_2$ \citep{ff89,mf91,kt10}.

\begin{figure*}
\includegraphics[angle=90,scale=.70]{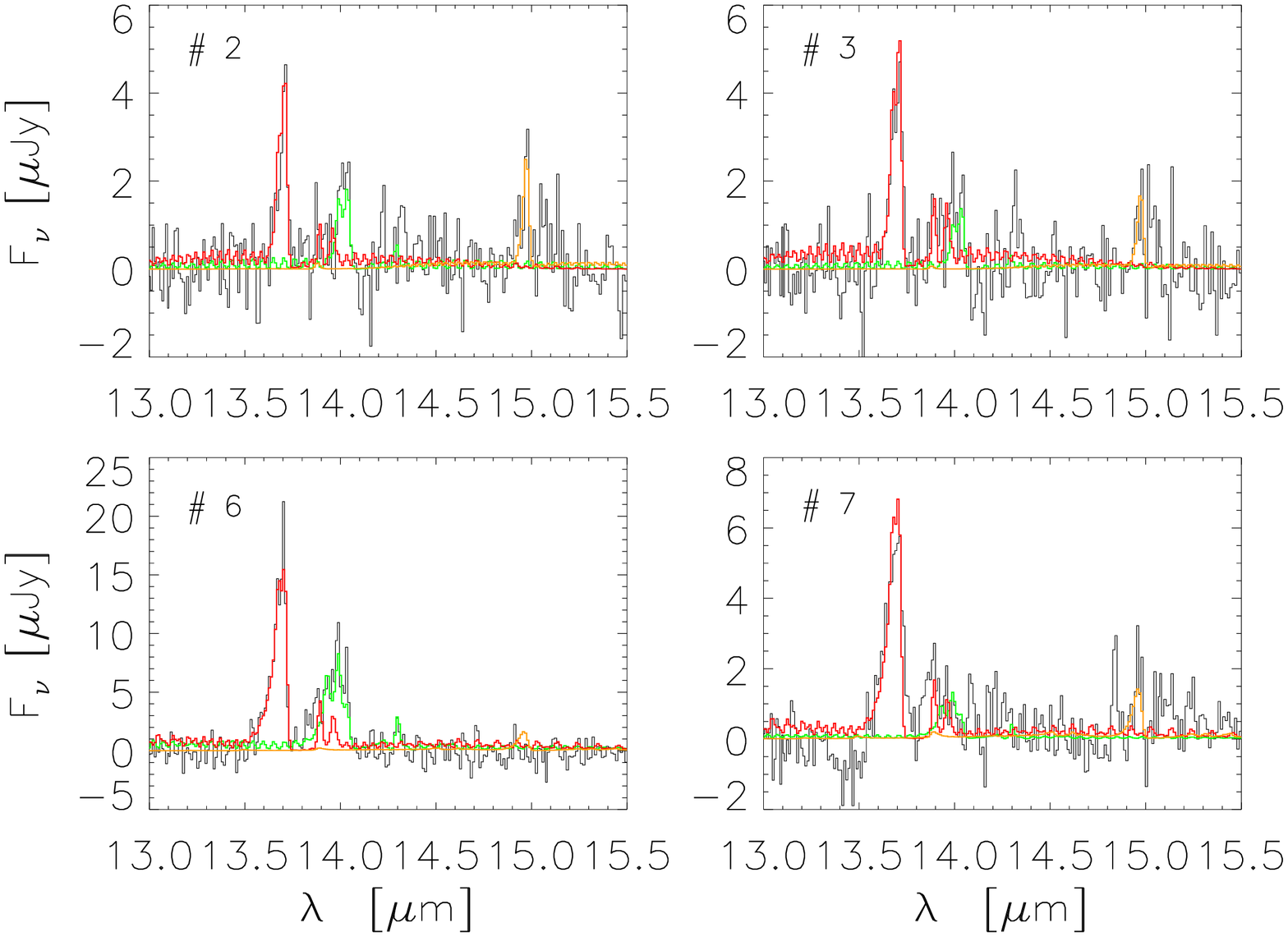}
\caption{Continuum-subtracted spectra (black lines) of the four very-low mass stars/brown dwarfs where we detect molecular rovibrational emission bands. Best fit synthetic spectra of C$_2$H$_2$ (red histogram), HCN (green histogram), and CO$_2$ (orange histogram) are overplotted on the data. The best fit parameters are summarized in Table~\ref{table:resmolfit}.}
\label{fig:moldet}
\end{figure*}

Our approach shows that there are large uncertainties in determining the gas properties of even the strongest rovibrational bands, those from C$_2$H$_2$. Therefore we do not attempt to determine independently the gas properties for HCN, CO$_2$, and H$_2$O but rather assume the same temperature and emitting area as that found for C$_2$H$_2$. We then scale the column densities to match the integrated flux (or 3$\sigma$ upper limits) over the other molecular bands. In the case of HCN, we first subtract the best fit C$_2$H$_2$ model spectrum reported in Table~\ref{table:resmolfit} and recompute the HCN line flux. Best fits parameters are reported in Table~\ref{table:resmolfit} and the resulting synthetic spectra are shown in Fig.~\ref{fig:moldet}. For water we report column density upper limits for two temperatures: i) the temperature obtained from the best fit to the C$_2$H$_2$ band (third column in Table~\ref{table:resmolfit}) and ii) the temperature of 600\,K that is found to be typical to the water-emitting layer of T~Tauri disks \citep{carr11}. The emitting radius is always fixed to that found from the best-fit to the C$_2$H$_2$ band (fourth column in Table~\ref{table:resmolfit}). The column density of water is poorly constrained by our data and very sensitive to the assumed temperature. In particular, if water vapor is as cold as C$_2$H$_2$ in objects \#2 and 3 our line flux upper limits are not stringent enough to place any useful constraint on the amount of water on the surface of these disks. This is because in offsetting the low flux due to the cold gas water lines become optically thick and grow very slowly with increasing column density. As an example for $T_{\rm H_2O}=240$\,K and $N_{\rm H_2O}=10^{20}$\,cm$^{-2}$ the 17.22\,\micron{} is the brightest of the water lines but has an optical depth at the line center of $\sim$50. Such cold water emission is not typical to T~Tauri disks \citep{carr11,salyk11}. However, lacking constraints from disks around lower-mass stars, we include the results from this extreme temperature in our discussion.

\section{Discussion}

\subsection{On the origin of the \neii{} emission toward BD disks}\label{discuss:neii}
Ground-based observations have spectrally resolved the brightest \neii{} lines detected with {\it Spitzer} toward T~Tauri stars \citep{herczeg07,boekel09,najita09,pascuccisterzik09,pascucci11,sacco12,baldovin12}. These data have demonstrated that \neii{} emission can either trace jets/outflows far from the star (typically toward high-accreting stars) or the hot ($\sim$10,000-5,000\,K) disk surface ionized by stellar EUV and/or X-ray photons (in moderate- to low-accreting stars, $M_{\rm acc} \lesssim 10^{-8}$\,M$_\sun$/yr). 
When disk emission is detected most \neii{} lines are blueshifted by a few km/s with respect to the stellar velocity pointing to unbound gas in a photoevaporative thermal wind (e.g. \citealt{pascuccisterzik09,sacco12}). t is debated whether \neii{} emission traces the EUV fully ionized or the X-ray partially ionized disk layer, hence mass loss rates via photo evaporation are yet unconstrained (e.g. the recent review by \citealt{alexander13}).

\begin{figure*}
\includegraphics[angle=90,scale=.70]{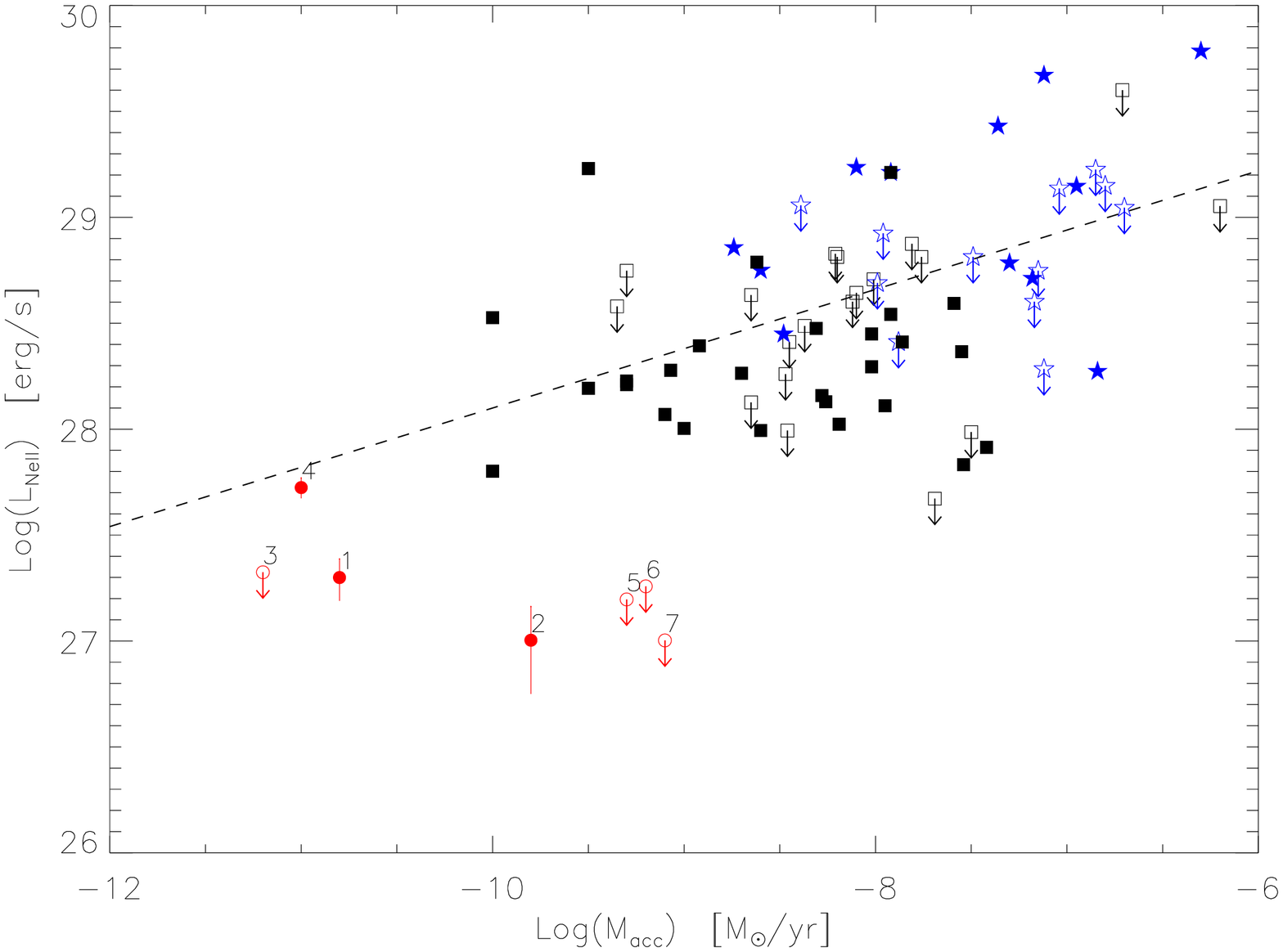}
\caption{\neii{} line luminosities versus mass accretion rates. 
Brown dwarf disks from this work are represented with red circles.
Black squares refer to T~Tauri disks (including transition disks) without jets, 
 and blue stars to jet sources from the literature 
\citep{guedel10,baldovin12,espaillat13}. 
Filled and open symbols are used for detections and non-detections, respectively. 
Our object \#8 does not have a measured mass accretion rate, hence it is not included in the figure.
}
\label{fig:neiimdot}
\end{figure*}

To decide whether our \neii{} detections trace jets/outflows or the BD disk surface, we plot the \neii{} luminosity ($L_{\rm{NeII}}$) as a function of the stellar mass accretion rate ($M_{\rm acc}$). Together with our sample, we include all other {\it Spitzer} detections and upper limits reported in the literature for T~Tauri stars  (see Fig.~\ref{fig:neiimdot}). The complete \neii{} sample spans almost six orders of magnitude in $M_{\rm acc}$ and three orders of magnitude in  $L_{\rm{NeII}}$. When we extrapolate the $L_{\rm{NeII}}$--$M_{\rm acc}$ relation found for T~Tauri stars with known jets \citep{guedel10} to the BD regime, we see that only the strongest of our \neii{} detections, that from source \#4, can be attributed to jets/outflows. Indeed, \citet{bouy08} detected strong forbidden optical lines (O~I, S~II, and N~II) in the high-resolution spectrum of source \#4 (2MASS~J04442713+2512164) suggesting that this object is undergoing significant mass loss via a jet/outflow. We conclude that most of the \neii{} emission from source \#4 is not associated with the disk but produced in circumstellar material shocked by jets/outflows.

BDs \#1 and 2 have lower $L_{\rm{NeII}}$ than source \#4,  by factors of $\sim$3 and 5 respectively (the detection toward \#2 is only tentative). It is interesting to speculate what X-ray/EUV fluxes are needed to reproduce these lower luminosities assuming that  the \neii{} emission originates entirely in the ionized surface of BD disks.
In the simplest scenario in which neon atoms are ionized solely by stellar EUV photons, we can convert the observed \neii{} luminosities into the rate of ionizing photons reaching the disk surface ($\Phi_{\rm EUV}$, HG09). The \neii{} detections from \#1 and 2 correspond to $\Phi_{\rm EUV}$ of 0.6-1.2$\times 10^{40}$\,s$^{-1}$. These values are $\sim$5-10 times lower than the $\Phi_{\rm EUV}$ inferred for T~Tauri stars \citep{alexander05,pascucci12} and represent  upper limits to the EUV luminosities impinging on the disk surface. This is because part or most of the \neii{} emission could arise from a disk layer ionized by stellar X-rays, see below. It is unclear whether EUV photons are from the stellar chromosphere and/or accretion processes (see e.g. \citealt{alexander05}). If the latter dominate then $\Phi_{\rm EUV}$ should scale with the accretion luminosity or equivalently with the stellar FUV luminosity ($L_{\rm FUV}$), since the two quantities are tightly correlated \citep{yang12}. $L_{\rm FUV}$  for T~Tauri stars cluster between $0.001-0.01$\,L$_\odot$ (see Fig.~12 in \citealt{yang12}) while the $L_{\rm FUV}$ we estimate for our sample are $\sim 10-1,000$ times lower (see Tables~\ref{table:propt} and \ref{table:inferred}). X-ray luminosities ($L_{\rm X}$) decrease with stellar mass less steeply than $L_{\rm FUV}$. The $L_{\rm X}$ of our sources (most of which are estimated from the source mass, see Table~\ref{table:inferred}) are $\sim 10-100$ times lower than the average $L_{\rm X}$ of T~Tauri stars ($\sim 10^{30}$\,erg/s, \citealt{guedel07}). These trends suggest that X-rays rather than EUV photons are the main ionizing source of neon atoms in BD disk atmospheres. 
According to the models of \citet{schisano10}, unscreened X-ray luminosities of $\sim 5 \times 10^{28}$\,erg/s are necessary to  produce $L_{\rm{NeII}}\sim 10^{27}$\,erg/s, which is close to our \neii{} fluxes and 3$\sigma$ upper limits. Therefore, the fact that  the majority of our disks remain undetected in the \neii{} line is consistent with their low X-ray luminosities. 
In summary, a disk origin is plausible for the \neii{} emission reported toward BDs  \#1 and 2 and X-rays ionization is likely the major contributor to neon ions in their disk atmosphere.

The detection of \neii{} emission from BD disk atmospheres shows that even young very low-mass objects have enough high-energy photons to create a fully or partially ionized hot surface. Whether this surface is unbound and part of a photoevaporative flow, as in moderate- to low-accreting T~Tauri stars, cannot be tested with the IRS low resolving power. A resolution of $\sim10$\,km/s is needed to spectrally resolve lines tracing the flow and detect the few km/s blueshifts in peak centroid with respect to photospheric lines. Due to the faintness of our sample, and BD disks in general, this test will have to wait for high-resolution spectrographs mounted on 40\,m-class telescopes.

\subsection{Warm gas masses from H$_2$ detections}\label{discuss:warmh2}

H$_2$ pure rotational lines are found to be relatively weak toward T~Tauri stars with respect to emission from other molecules and from atoms/ions. The {\it Spitzer} "Cores to Disks" legacy program reports that only 6 out of 76 young ($\sim$Myr-old) disks present the H$_2$ S(2) line in their mid-infrared spectra \citep{lahuis07}. The H$_2$ S(1) line is detected in only 1 out of 76 spectra. Because of the large beam of {\it Spitzer} and the lack of 'sky' observations for this program, extended (envelope, local cloud, jet) emission can potentially contaminate some of the reported line fluxes \citep{lahuis07}. Deeper observations and a better sky subtraction increase the H$_2$ detection rate. \cite{carr11} identify the H$_2$ S(1) transition toward 6 out of their 11 Taurus T~Tauri stars but do not provide H$_2$ S(2) fluxes because of contamination from much stronger H$_2$O and OH lines around 12.28\,\micron . Our observing technique and data reduction are similar to theirs. In addition, our sample sports weaker water and OH emission than T~Tauri disks (see Sect.~\ref{sect:molecules}) resulting in an H$_2$ detection rate similar to that reported by \cite{carr11} even toward sources that are about an order of magnitude fainter in the continuum.

Similarly to the continuum emission, our H$_2$ fluxes are about an order of magnitude lower than those observed toward T~Tauri disks \citep{carr11} and predicted from disks irradiated by FUV and X-ray photons from T~Tauri stars \citep{nomura07}. This is in line with the lower FUV and X-ray luminosity of very low-mass stars and BDs and suggests that our H$_2$ lines are indeed tracing disk gas.
In the assumption that this gas is in LTE and lines are optically thin\footnote{this is a good approximation because of the small A-values of the pure rotational transitions}, we can use the H$_2$ S(2) and  H$_2$ S(1) detections (or upper limits in one of the two) to compute a gas excitation temperature $T_{\rm ex}$  (see e.g. \citealt{thi01} for the equations). Our measurements are consistent with $F_{\rm H_2 S(1)} \sim F_{\rm H_2 S(2)}$ which implies $T_{\rm ex}$ = 645\,K with the assumptions outlined above. Including the 1$\sigma$ uncertainties on the measured fluxes for objects \#4 (Table~\ref{table:neiih2}) expands the range of possible $T_{\rm ex}$ to $450-1,200$\,K.  If we then use these temperatures to compute a gas mass, we find that the H$_2$ rotational lines trace only $\sim2-7 \times 10^{-4}$\,M$_{\rm J}$ (with the largest value for gas at the lowest $T_{\rm ex}$ in the range reported above). Gas masses of BD disks ($M_{\rm disk}$) have not been measured. However, the few sub-mm and mm detections of the continuum dust emission suggest that the $M_{\rm disk}/M_{\rm BD}$ is a few percent, if the gas-to-dust-ratio is equal to the ISM value of 100 \citep{klein03,scholz06,ricci12}. This result points to total BD disk masses of $\sim$M$_{\rm J}$, more than three orders of magnitude larger than what we find from the H$_2$ pure rotational lines. This discrepancy is not surprising since models of disks irradiated by stellar FUV and X-rays predict that H$_2$ pure rotational lines trace mainly the warm surface of protoplanetary disks and are thus insensitive to the more abundant and cooler disk gas (e.g., \citealt{nomura07}). 

The H$_2$ and \neii{} results combined show that BD disks have overall a structure similar to T~Tauri disks: a hot ionized layer tracing the uppermost disk atmosphere and a warm ($\sim$600\,K) molecular layer at lower vertical heights.

\subsection{An enhanced carbon chemistry in BD disks?}\label{discussion:highC}

We use the molecular column densities inferred from our LTE modeling (Sect.~\ref{sect:let}) to compute relative abundances of simple organic molecules and water in disks around very low-mass stars and BDs. This approach is complementary to the discussion of line flux ratios in Sect.~\ref{sect:molecules} and has been previously taken for T~Tauri disks \citep{carr11,salyk11}. 
There are a number of caveats in using this simple approach (e.g., \citealt{carr11}). Perhaps the most important one is that the radial and vertical regions probed by these molecules may be different while they are often assumed the same due to the lack of spatial information with {\it Spitzer} observations. In addition, these infrared emission lines trace the warm disk surface which may not be representative of the overall disk composition.

\begin{figure*}
\includegraphics[scale=1.]{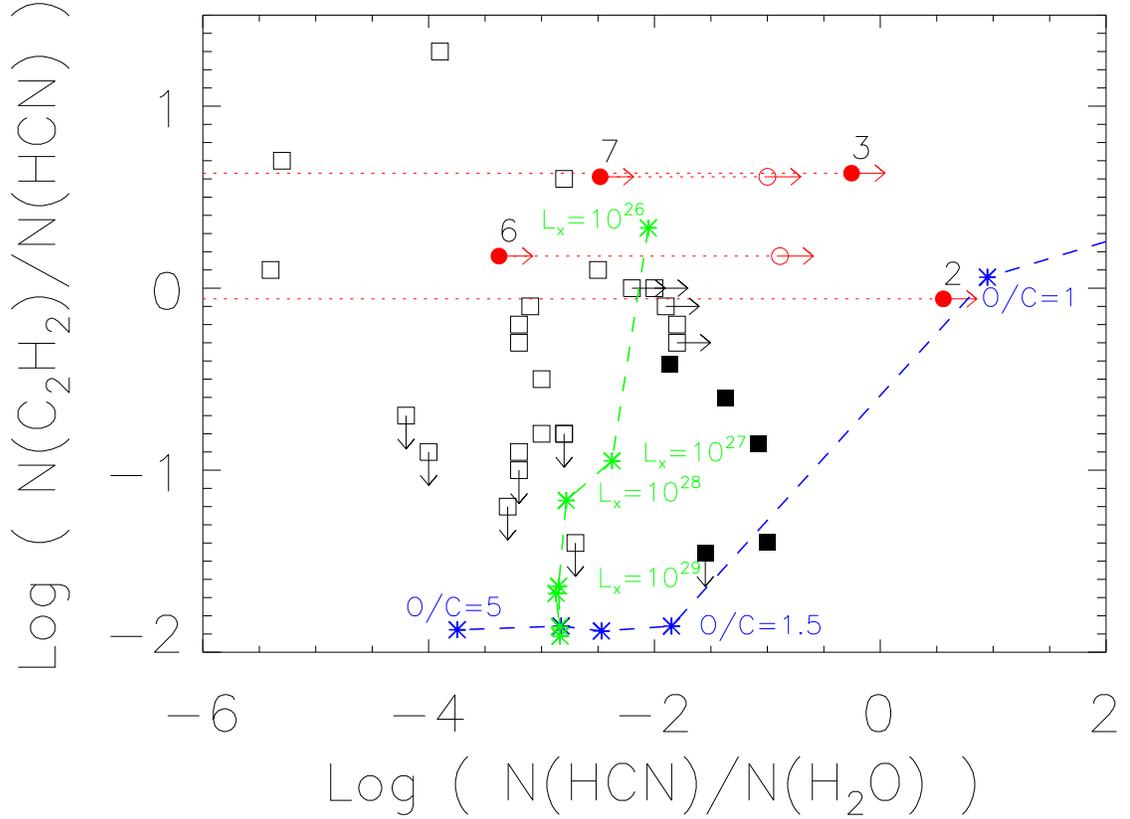}
\caption{Column density of C$_2$H$_2$ to HCN plotted versus that of HCN to H$_2$O.
Red circles are brown dwarf disks from this work: filled and empty symbols are for the two closest temperatures to the $T_{\rm{C_2H_2}}$ in Table~\ref{table:resmolfit} (400 and 600\,K for \#2 and 3, 600 and 960\,K for \#6 and 7).
Emitting radii are also listed in Table~\ref{table:resmolfit} and are assumed the same for all molecules.
Black squares are T~Tauri disks from \citet{carr11} while empty squares are T~Tauri disks from \citet{salyk11}.
Green and blue asterisks joined by dashed lines are model predictions from \citet{najita11} for different $L_{\rm X}$ and O/C ratios respectively (their Tables~4 and 5). The current data hint to brown dwarf disk atmospheres having a higher carbon-to-oxygen ratio than T~Tauri disk atmospheres.  
}
\label{fig:gridmol}
\end{figure*}

With these caveats in mind, our modeling results suggest that C$_2$H$_2$ is typically more abundant than HCN in BD disks and both C$_2$H$_2$ and HCN are more abundant than CO$_2$. The column density of water is very sensitive to the assumed gas temperature and useful constraints on its abundance can be placed only if gas is at $T\ge 600$\,K even in objects \#2 and 3. Figure~\ref{fig:gridmol} compares column density ratios of C$_2$H$_2$/HCN versus  HCN/H$_2$O for our BD disks (red symbols) and T~Tauri disks from the literature (black symbols). Horizontal red dashed lines connect the two water column densities reported in Table~\ref{table:resmolfit}. The most obvious difference between BD and T~Tauri disks is in their C$_2$H$_2$/HCN column density ratios. BD disks appear to have a systematically higher C$_2$H$_2$/HCN column density ratio: The median ratio for BD disks is $\sim$4 while that for the complete T~Tauri sample is $\sim 0.6$, not considering the seven C$_2$H$_2$ upper limits in the T~Tauri sample. Thus, on average
C$_2$H$_2$ is about seven times more abundant than HCN in BD disks than in T~Tauri disks. This difference was first pointed out from the analysis of low-resolution {\it Spitzer} spectra \citep{pascucci09} and is corroborated by our higher resolution spectra on a different sample of BD disks. This finding is also in line with the typically higher C$_2$H$_2$/HCN strength in BD than in T~Tauri disks (Sect.~\ref{sect:molecules} and Fig.~\ref{fig:fluxratios}).
If water vapor is warm $T \ge 600$\,K, Figure~\ref{fig:gridmol} also hints to a reduced abundance of H$_2$O in BD disks with respect to T~Tauri disks. This possible trend is also apparent in the model-independent plot of line flux ratios, especially for the U~Sco sources (Fig.~\ref{fig:fluxratios}). 

We now turn to theoretical models of disks irradiated by stellar FUV and X-rays  \citep{agundez08,najita11,walsh12} to seek explanations for the trends in molecular abundances discussed above. 
We note that models report vertical column densities integrated over the height of the disk while observations measure line-of-sight column densities and infrared observations like ours only trace the warm disk atmosphere. So any comparison (in this as well as in other works) relies on the assumption that vertically integrated column densities are representative of the column densities in the warm disk atmosphere.

The models of \cite{agundez08} are for typical T~Tauri disks around Sun-like stars and include FUV heating but not X-ray heating. They find that the steady state vertical column density of H$_2$O is relative constant out to $\sim$3\,AU from the star. On the contrary, the organic molecules HCN, C$_2$H$_2$, and CO$_2$ have a stronger radial dependence, with  C$_2$H$_2$ (and even more CO$_2$) becoming less abundant in the very inner disk (inside $\sim$0.5\,AU).
At $\sim$1\,AU, the radius likely probed with {\it Spitzer} around T~Tauri stars,
HCN is more abundant than C$_2$H$_2$ by a factor of $\sim 10$, in agreement with
inferred column density ratios for several T~Tauri disks \citep{carr11,salyk11}. 
Only beyond $\sim$2.5\,AU C$_2$H$_2$ becomes more abundant than HCN in the \cite{agundez08} model. 
The models of \cite{walsh12} include both FUV and X-ray heating. As in \cite{agundez08} they predict that H$_2$O is the most abundant molecule in the gas phase (after H$_2$) inside the snow line. However,  unlike \cite{agundez08}, they find that HCN remains more abundant than C$_2$H$_2$ throughout the all radial extent of the disk. The differences among these model results are likely due to differences in the computed gas disk temperature. Our higher C$_2$H$_2$/HCN in BD disks cannot be explained as the result of these two molecules tracing different radii in the \cite{walsh12} models. In the \cite{agundez08} models, our result would suggest that C$_2$H$_2$ traces cooler gas, at larger radii, than HCN. This does not seem to be case since our $T$ estimates from fitting the C$_2$H$_2$ band do a good job in reproducing the profile of the HCN emission band.
If the predicted vertical column densities reflect atmospheric column densities, a depletion of water in BD disks, as hinted by our line flux ratios and LTE modeling, cannot be explained as the result of radial dependencies in molecular column densities.
Detailed thermo-chemical models of BD disks should be carried out to determine whether the different properties of BD and their disks could explain the observed trends.


Along these lines, \cite{najita11} have explored how changing the properties T~Tauri disks and stellar heating might affect molecular abundances. As a cautionary note the models of 
\cite{najita11} include accretional and stellar X-ray heating but neglect FUV heating and UV-induced processes, which generally are found to be more important than X-ray processes \citep{walsh12}.
\cite{najita11}  find that low accretional heating (lower disk temperature) reduces the warm molecular columns of most species but further decreases the C$_2$H$_2$/HCN ratio while keeping large column densities of H$_2$O. Both trends contrast with our results suggesting that the lower accretional heating of BD disks is not responsible for the observed trends in molecular abundances. Increasing the grain size from 0.7 to 7\,\micron{} in the disk atmosphere increases the warm molecular columns of most species as well as the C$_2$H$_2$/HCN ratio. While this trend is in qualitative agreement with the observed one, 7\,\micron{} grains in the disk atmosphere enhance the C$_2$H$_2$/HCN ratio to only 0.07, still too low in comparison with the observed one. At the same time it is very unlikely that BD disks have even larger grains in their atmosphere because many of them show 10\,\micron{} silicate emission features in their low-resolution {\it Spitzer} spectra \citep{pascucci09}. Such emission features point to the presence of dust grains with sizes $\le$5\,\micron. Another way of increasing the C$_2$H$_2$/HCN ratio in the models of  \cite{najita11} is by decreasing $L_{\rm X}$ (green symbols in Figure~\ref{fig:gridmol}). However, the C$_2$H$_2$/HCN ratios we measure for BDs are reproduced for an $L_{\rm X}$ of $10^{26}$\,erg/s, almost two orders of magnitude lower than that in our sample (see Tables~\ref{table:propt} and \ref{table:inferred}). A decrease in $L_{\rm X}$ should also result in a very modest change (just a factor of a few) in the HCN/H$_2$O ratio, with water being more abundant than HCN by a factor of $\sim 100$ like in T~Tauri disks. This latter trend is not consistent with our BD sample if the water-emitting layer is warm ($T\ge600$\,K).
Even if the water non-detections alone cannot exclude a lower gas temperature, the detection of H$_2$ and other organic molecules show that a warm $T\sim600$\,K molecular layer is present in BD disk atmospheres. In T~Tauri disk atmospheres this layer has also a large H$_2$O column density inside $\sim$1\,AU and produces the observed water emission spectrum \citep{najita11}. Hence, it is plausible that there is some warm ($T\ge600$\,K) water in BD disk atmospheres but perhaps not enough to be detectable in our spectra.
Finally, lowering the O/C ratio from 1.5 to 1 (blue symbols in Figure~\ref{fig:gridmol}) enhances the C$_2$H$_2$/HCN and HCN/H$_2$O molecular ratios, which is in line with our trends. In the framework of these models, BD disk atmospheres appear to have a lower O/C ratio, close to 1, than T~Tauri disk atmospheres (see also \citealt{mandell12} who find an O/C=1.5 from the analysis of near-infrared C$_2$H$_2$ and HCN rovibrational bands in T~Tauri disks). When oxygen is depleted relative to carbon water abundances decrease while the abundance of hydrocarbons, especially C$_2$H$_2$ and CH$_4$ increase \citep{najita11}. Unfortunately, the high-resolution {\it Spitzer} spectrograph does not cover the rovibrational band of CH$_4$ and testing the inference of an enhanced carbon chemistry in BD disks will require additional observations. We also note that the CO$_2$ abundance decreases only slightly (by about a factor of 2) when going from a O/C=1.5 to 1 meaning that C$_2$H$_2$ and HCN remain more abundant than CO$_2$, in agreement with our finding.

Finally, we wish to discuss our results in the context of the proposed correlation between the strength of HCN/H$_2$O and the disk mass of T~Tauri stars. \cite{najita13} attribute this correlation to  higher disk masses producing more non-migrating planetesimals that lock up water ice beyond the snow line, thereby decreasing the O/C ratio in the inner disk. Of our very low-mass star and BD disks only three (\#1, 3, and 4) have measured masses \citep{scholz06}. Their average disk mass is $\sim 10^{-3}$M$_\odot$, which agrees with the disk mass over the stellar mass being about 1\% throughout the M spectral type and down to the brown dwarf regime \citep{andrews13,mohanty13}. If BDs would follow the trend proposed by \citet{najita13}, we should have measured HCN/H$_2$O flux ratios as low as $\sim$0.6 for the water line at 17.22\micron\footnote{Reading straight from Fig.~3 of \citet{najita13} the HCN/H$_2$O flux ratio is $\sim$0.2 but the water flux there is measured for the 17\micron{} group, which includes three water lines (17.12, 17.22, and 17.36\,\micron). Our HCN/H$_2$O flux ratio is computed from the 17.22\micron{} H$_2$O line only, which contributes to about a third of the 17\,\micron{} group flux.}
, indicating the prevalence of water vapor (high O/C ratio) in the inner region of low-mass disks. On the contrary, we find HCN/H$_2$O flux ratios all larger than 3 (see Fig.~\ref{fig:fluxratios}),  pointing to a depletion of water vapor in the inner regions of BD disks. 
Clearly, disks around very low-mass stars and BDs do not follow the proposed trend between the HCN/H$_2$O flux ratio and the disk mass.

Could our observations still be explained with an analogous process, i.e. could low-mass BD disks have formed more non-migrating icy planetesimals than T~Tauri disks? As we mention in the Introduction our BD sample may be slightly biased toward older objects with no strong jets/outflows. These disks may have had more time to produce large non-migrating icy planetesimals than the perhaps younger sample of T~Tauri disks discussed in the literature so far. Even for disks in the same star-forming region there is growing observational evidence that the first steps of planet formation proceeds faster in BD than in T~Tauri disks at radial distances tracing the same disk temperature. First, 10 and 20\,\micron{} silicate emission features suggest that the warm (100-500\,K, \citealt{attila09}) atmosphere of BD disks contains on average larger grains than that of T~Tauri disks \citep{pascucci09,riaz12}. Second, detailed SED modeling of 'coeval' disks finds that disks around very low-mass stars and BDs are more settled than disks around T~Tauri stars, which also points to faster grain growth followed by dust settling \citep{szucs10}. Finally, the detection of millimeter-sized grains in two young BD disks \citep{ricci13} demonstrates that even in the cold outer regions of these low-mass disks large grains can form in a Myr timescale and be retained. The recent theoretical work of \citet{pinilla13} highlights some of the main physical processes that could  contribute to the observed differences between BD and T~Tauri disks. 
\citet{pinilla13} show that the growth timescale of micron-size particles due to settling is directly proportional to the square root  of the stellar luminosity. This implies that small grains in BD disks grow an order of magnitude faster than those in T~Tauri disks that trace regions at similar temperatures. At the same time, because the radial drift velocity is inversely proportional to the square root of the stellar mass, dust particles in BD disks are depleted faster than in T~Tauri disks. While a mechanism to slow down the depletion of millimeter/meter-size grains is needed in T~Tauri disks \citep{pinilla12}, that mechanism must be more efficient in BD disks to explain the recent detections of mm dust particles. 
The models of \citet{pinilla13} support the view that the first steps of planet formation are faster in BD than in T~Tauri disks. Unfortunately, these models do not follow the growth of dust up to planetesimals ($\sim$km in size), which is when gas drag does not significantly affect their orbital dynamics. Our inference of a lower O/C in BD disks than in T~Tauri disks perhaps suggests that the faster dust growth in BD disks extends to the stage of non-migrating planetesimals.

\section{Summary}
We analyzed a unique dataset of high-resolution {\it Spitzer} spectra around eight very low-mass star and brown dwarf disks. We report the first detections of Ne$^+$, H$_2$, CO$_2$, and tentative detections of H$_2$O toward these very faint and low-mass disks. Our main results can be summarized as follows:
\begin{itemize}
\item Two of our \neii{} 12.8\,\micron{} detections are likely tracing the hot ($5,000-10,000$\,K) surface of BD disks. \neii{} line fluxes and upper limits, in combination with ancillary BD properties, suggest that X-rays rather than EUV are the main ionizing photons of BD disk atmospheres
\item The H$_2$ S(2) and S(1) fluxes are consistent with an origin in the warm ($T\sim600$\,K) surface of BD disks. Hence, as in T~Tauri disks, mid-infrared H$_2$ lines do no probe most of the disk mass
\item The C$_2$H$_2$/HCN flux and column density ratios are on average higher for BD than for T~Tauri disks, confirming the previous trend found in low-resolution {\it Spitzer} spectra \citep{pascucci09}
\item The HNC/H$_2$O flux ratios are, on average, also higher for BD than for T~Tauri disks. Our LTE modeling extends this result to column density ratios if mid-infrared H$_2$O lines trace warm ($T \ge 600$\,K) gas.
\end{itemize}
While the larger C$_2$H$_2$/HCN ratio is well established when combining this and previous results, the larger HNC/H$_2$O column density ratio in BD disks is still tentative. Future observations with JWST/MIRI will be easily able to followup such faint disks and detect weak water lines or place much tighter constraints on the amount of water than what we could do with {\it Spitzer}/IRS.

We find that higher C$_2$H$_2$/HCN {\it and} HNC/H$_2$O ratios can be explained by disk models with a reduced O/C ratio in the inner disk. This low ratio may be linked to the formation of non-migrating planetesimals locking up water beyond the snow line \citep{najita11}. In this scenario BD 'coeval' to  T~Tauri disks should be able to form more efficiently non-migrating icy planetesimals. Dust observations and theoretical models suggest that the first steps of planet formation, up to the growth of mm-size grains, proceed faster in BD than in T~Tauri disks. 
The observations we present here hint that this faster growth may extend to km-size bodies. 

The main implication of this scenario is that the inner regions of BD disks are overall 'drier' than those of T~Tauri disks. With a O/C ratio close to 1, the bulk composition of rocky planets around BDs should include large amounts of C phases \citep{bond10} and be overall different from any terrestrial planet in our Solar System.

\acknowledgments
This work is based on observations made with the Spitzer Space Telescope, which is operated by the Jet Propulsion Laboratory, California Institute of Technology under a contract with NASA. I.P. acknowledges support from the NASA/ADP Grant NNX10AD62G. Basic research in infrared astrophysics at the Naval Research Laboratory is supported by 6.1 base funding.



{\it Facilities:} \facility{Spitzer}, \facility{2MASS}.



\appendix

\section{Broad-band features in the spectrum of 2MASS~J16053215-1933159}\label{obj7}
\begin{figure}
\includegraphics[scale=1.]{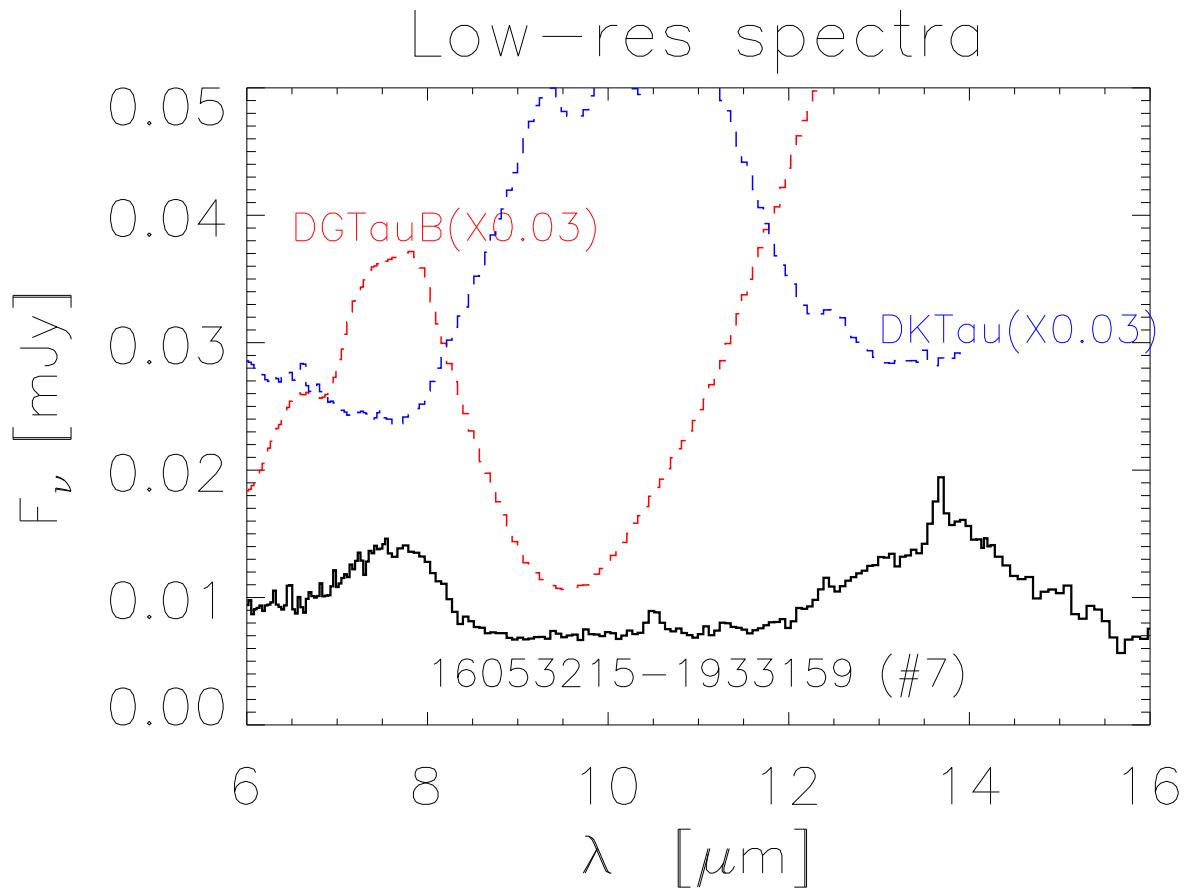}
\caption{IRS low-resolution spectrum of 2MASS~J16053215-1933159 (\#7, solid black line). Red dashed line is the low-resolution spectrum of DGTauB scaled by the factor in parenthesis. The low-resolution spectrum of DKTau, scaled by the factor in parenthesis, is shown with a blu dashed line. Object~7 appears to have a shallow 10\,\micron{} silicate absorption feature indicative of a relatively high disk inclination and predominantly micron-size dust grains.  
}
\label{fig:obj7}
\end{figure}

The IRS low-resolution spectrum of 2MASS~J16053215-1933159 (\#7) is shown in Fig.~\ref{fig:obj7}. Two broad-band features are present in this spectrum: one peaking around 7.5\,\micron, the other around 14\,\micron{} (this latter feature is also clearly detected in the IRS high-resolution spectrum, Fig.~\ref{fig:highresspk}). We have inspected the combined Nod1 and 2 images and found no evidence for extended emission at these wavelengths beyond the 3\farcs6 slit width. Hence, any observed emission is associated to the star/disk system. We have also searched the literature for known broad-band dust emission features at these wavelengths but found none peaking at 7.5 and/or 14\,\micron. One possibility is that we are seeing this disk at relatively high inclination such that the prominent 10\,\micron{} silicate feature is seen in absorption, producing two apparent emission features shortward and longward of 10\,\micron. We have re-reduced many of the Taurus low-resolution spectra of disks available in the Spitzer Heritage Archive and found that the spectrum of the highly inclined disk (75$^\circ$, \citealt{padgett99}) DG~TauB is the one that best resembles that of object 7 (see Fig.~\ref{fig:obj7} red dashed line, AOR: 25680128). The 10\micron{} absorption feature of DG~TauB is much narrower and deeper than that of object~7 reflecting the predominance of small/sub-micron size dust grains  (see also \citealt{kruger11}). Fig.~\ref{fig:obj7} also shows the IRS low-res spectrum of DK~Tau (AOR: 3531264), a T~Tauri disk with a prominent but broad 10\,\micron{} silicate emission feature indicative of micron-size dust grains in its atmosphere. Like DK~Tau the disk atmosphere of 2MASS~J16053215-1933159 appears to be dominated by micron-size grains.

\end{document}